\documentclass{aa}  

\usepackage{graphicx}
\usepackage{txfonts}
\usepackage{lipsum}
\usepackage{caption}
\usepackage{subcaption}
\usepackage{lscape}
\usepackage{placeins}

\usepackage[
 bookmarks=true,
 pdfnewwindow=true,
 colorlinks=true,
 linkcolor=blue,
 citecolor=blue,
 filecolor=blue,
 urlcolor=blue,
 final=true,
]{hyperref}

\usepackage[
 print-unity-mantissa=false,
 exponent-product=\ensuremath{\cdot},
 uncertainty-mode = separate,
]{siunitx}

\usepackage{amsmath}
\usepackage{booktabs}
\usepackage{makecell}
\usepackage{xcolor}

\DeclareSIUnit{\photon}{ph}
\DeclareSIUnit{\erg}{erg}
\DeclareSIUnit{\deg}{deg}
\DeclareSIUnit{\gauss}{\text{\ensuremath{\textnormal{G}}}}
\DeclareSIUnit{\solarmass}{\text{\ensuremath{\textnormal{M}_\odot}}}
\DeclareSIUnit{\solarradius}{\text{\ensuremath{\textnormal{R}_\odot}}}
\DeclareSIUnit{\year}{yr}
\DeclareSIUnit{\parsec}{pc}
\DeclareSIUnit{\magnitude}{\text{\ensuremath{\textnormal{mag}}}}

\def\asrc {AGL~J1736$-$3250}
\def\isrc {IGR~J17354$-$3255}

\def \swift {{\it Swift}}
\def \xmm {{\it XMM-Newton}}

\def \arcsec {\hbox{$^{\prime\prime}$}}
\def \gray {$\gamma$-ray }
\def \grays {$\gamma$-rays }

\definecolor{airforceblue}{rgb}{0.36, 0.54, 0.66}

\begin{document}

\title{AGILE detection of transient \gray emission from the region of the supergiant fast X-ray transient source IGR~J17354$-$3255}
\titlerunning{AGILE detection of transient emission from the IGR~J17354$-$3255 region}

\author{
    Andrea Bulgarelli\inst{1}\fnmsep\thanks{E-mail: \href{mailto:andrea.bulgarelli@inaf.it}{andrea.bulgarelli@inaf.it}}
    \and Gabriele Panebianco\inst{1}
    \and Vito Sguera\inst{1}
    \and Marco Tavani\inst{2,3}
    \and Valentina Fioretti\inst{1}
    \and Ambra di Piano\inst{1}
    \and Nicolò Parmiggiani\inst{1}
    \and Patrizia Romano\inst{4}
    \and Stefano Vercellone\inst{4}
    \and Alessio Aboudan\inst{5}
    \and Francesco Longo\inst{6,7}
    \and Giacomo Principe\inst{6,7,14}
    \and Francesco De Palma\inst{8,9}
    \and Emanuele Dolera\inst{10}
    \and Carlotta Pittori\inst{11,12}
    \and Fabrizio Lucarelli\inst{11,12}
    \and Francesco Verrecchia\inst{11,12}
    \and Andrew W. Chen\inst{13}
    \and Angela Bazzano\inst{2}
}
\institute{
    INAF - Osservatorio di Astrofisica e Scienza dello spazio di Bologna (OAS), Via Piero Gobetti 93/3, 40129 Bologna, Italy
    \and INAF - Istituto di Astrofisica e Planetologia Spaziali (IAPS), Via del Fosso del Cavaliere 100, 00133 Roma, Italy
    \and Dipartimento di Fisica, Universitá di Roma Tor Vergata, via della Ricerca Scientifica 1, I-00133 Roma, Italy
    \and INAF -- Osservatorio Astronomico di Brera, Via E.\ Bianchi 46, I-23807, Merate, Italy
    \and CISAS G. Colombo University of Padua, Padua, Italy
    \and Dipartimento Fisica, Universitá di Trieste, via A. Valerio 2, I-34127 Trieste, Italy
    \and INFN Sezione di Trieste and Università degli Studi di Trieste, Via Valerio 2 I, 34127 Trieste, Italy
    \and Università del Salento, Dipartimento di Matematica e Fisica “E. De Giorgi”, Lecce, Italy
    \and INFN, Sezione di Lecce, Lecce, Italy
    \and Department of Mathematics, University of Pavia, Pavia, Italy
    \and ASI Space Science Data Center (SSDC), Via del Politecnico snc, I-00133 Roma, Italy
    \and INAF-Osservatorio Astronomico di Roma, Via di Frascati 33, I-00078 Monte Porzio Catone, Italy
    \and School of Physics, Wits University, Johannesburg, South Africa
    \and INAF -- Istituto di Radioastronomia (IRA), I-40129 Bologna, Italy
}
\authorrunning{A. Bulgarelli et al.}

\date{Received July 8, 2025; accepted April 13, 2026}

\abstract 
{On April 14, 2009, the AGILE satellite detected a \gray flare from an unknown transient source.
Subsequent X-ray follow-up observations with Swift and INTEGRAL identified the supergiant fast X-ray transient (SFXT) \isrc{} as the best candidate counterpart, based on positional coincidence and a similar temporal behaviour.
Aside from this hint, no SFXT has been firmly detected at high energies to date.
Overall, SFXTs comprise a subclass of high-mass X-ray binaries (HMXBs) that host a massive OB supergiant star as a companion donor.
They tend to display the most extreme X-ray variability among HMXBs.
These systems might be able to emit photons at MeV-TeV energies in the form of fast flares lasting from hours to a few days, with a low-duty cycle.}
{In this work, we analyse archival AGILE data to search for \gray flares consistent with \isrc{} and evaluate their possible physical origin.} 
{We identified a transient source, \asrc{}, which emitted $19$ \gray flares and was seen to be positionally consistent with \isrc{}.
Most flares, detected on a $\SI{1}{\day}$ timescale, concentrate most of their emission on two, four, and six hour timescales, resembling those observed in the X-ray band from \isrc{}.}
{An orbital phase analysis revealed that approximately half of the \gray activity occurs around the apastron passage of the compact object hosted in the binary system.
We also incorporated archival Swift and INTEGRAL observations to provide phase-folded light curves at lower energies.
Our collected results strongly support a physical association between \isrc{} and \asrc{}, offering evidence that SFXTs could constitute a new class of Galactic high-energy transient emitters.}
{}

\keywords{
    \href{http://astrothesaurus.org/uat/733}{High mass X-ray binary stars (733)}: IGR~J17354$-$3255
    -- \href{http://astrothesaurus.org/uat/1853}{Gamma-ray transient sources (1853)}
    -- \href{http://astrothesaurus.org/uat/1858}{Methods: Data Analysis (1858)}
    -- \href{http://astrothesaurus.org/uat/2109}{Time domain astronomy (2109)}
}

\maketitle
\nolinenumbers
\section{Introduction}
\label{sec:intro}

High-mass X-ray binaries (HMXBs) are binary stellar systems composed of a compact object that is either a neutron star or a black hole, accreting matter from a companion with a mass $\gtrsim \SI{10}{\solarmass}$, typically an O or B star.
X-ray emission in most HMXBs is accretion-powered, resulting from the fall of plasma towards the compact object, thereby releasing gravitational potential energy and heating up \citep[e.g. ][]{Kretschmar_HMXBs_2019, Fornasini_HMXBs_2023}.

Several HMXBs have recently been established as \gray emitters, providing evidence of efficient particle acceleration.
Some of them are classified as \gray binaries, while others are deemed to be $\gamma$-ray-emitting X-ray microquasars.
The nine gamma-ray binaries that are currently known emit the bulk of their luminosity at energies $\gtrsim \SI{1}{\mega\electronvolt}$ \citep[e.g. ][]{Chernyakova_gammaraybinaries_2019, bordas_GREB_2024}, in contrast to $\gamma$-ray-emitting X-ray microquasars, which peak in the X-ray band.
Gamma-ray binaries may either be `accretion-powered' microquasars or `rotation-powered' pulsars.
In the latter case, accretion is suppressed by a strong pulsar wind, and \grays are produced at the shock interface between the pulsar and companion winds \citep{Dubus_gammaraybinaries_2013, paredes_bordas_gammaraybinaries_2019}.

Over the past two decades, a new subclass of HMXBs has been discovered, the supergiant fast X-ray transient \citep[SFXTs; for a review see ][]{martinez_winds_2017}.
These systems host a compact object accreting material from a supergiant OB-type star and are characterised by short, sporadic X-ray outbursts lasting from a few hours to a few days.
Initially, SFXTs were discovered by INTEGRAL \citep{Sguera:2005, Sguera:2006, Neguerela:2006}.
Today they represent the most extreme case of X-ray variability among HMXBs, exhibiting: (1) flaring X-ray activity on timescales from a few minutes to a few hours; (2) very low duty cycles ($0.1-5\%$ above $\SI{20}{\kilo\electronvolt}$); and (3) dynamic ranges (of flux variations) up to five or six orders of magnitude below $10$ keV  \citep{Romano2015:17544sb,Sidoli_SFXT_XTEJ1739_2023}.
Approximately ten SFXTs are currently known and the number of candidates is approximately on the same level.
These sources display X-ray luminosities of $L_X \sim 10^{33-34}\,\si{\erg\per\second}$ in low-luminosity states and $L_X \sim 10^{36-37}\,\si{\erg\per\second}$ during flares, although quiescent luminosities of $L_X \sim \SI{E31}{\erg\per\second}$ and flares reaching $L_X \sim \SI{E38}{\erg\per\second}$ have occasionally been observed.
Some SFXTs show X-ray pulsations and spectral properties similar to other accreting pulsars in HMXBs; namely, an absorbed, flat power law below $\SI{10}{\kilo\electronvolt}$ (photon index between $0$ and $1$), with a high-energy cut-off at $\approx 10-30$~keV.
The physical mechanisms driving this peculiar phenomenology remain unclear and have been debated at length in the literature \citep[for a recent review see][]{Kretschmar_HMXBs_2019}.

An open question regarding SFXTs is whether they have the capacity to emit \grays in the range from $\si{\mega\electronvolt}$ to $\si{\tera\electronvolt}$ energies \citep{2009ApJ...697.1194S, Li_SFXTfromAntimagnetars_2011, sguera_IGRJ17354-3255_2013, 2025_CTAO_GalacticTransients}.
As in the case of \gray binaries and $\gamma$-ray-emitting X-ray microquasars, SFXTs might also produce high-energy (HE $\sim \si{\mega\electronvolt}-\si{\giga\electronvolt}$ range) and very-high-energy (VHE $\sim \si{\tera\electronvolt}$ range) emission, as these systems contain the same components; namely, a compact object and a massive early-type companion star.
However, detecting such emission is a non-trivial task for current HE and VHE instruments, as it likely consists of unpredictable flares with short duration, small duty cycles, and relatively low flux.
To date, several circumstantial pieces of evidence have suggested that SFXTs might be the counterparts to unidentified transient HE and VHE sources \citep{2009arXiv0902.0245S,Sguera:2011}.
Among these, \isrc{} stands out as one of the most promising cases.

\isrc{} is a transient hard X-ray object discovered by INTEGRAL \citep{Winkler_INTEGRAL_mission_2003} in 2006, in the framework of the Galactic Bulge monitoring program, during an outburst with a $20-60$~keV flux of $\sim \SI{2E-10}{\erg\per\second\per\centi\meter\squared}$ \citep{INT2006Tel, Kuulkers:2007}.
A comprehensive hard X-ray study performed with INTEGRAL in the $18-60$~keV energy band characterised \isrc{} as a weak, persistent, hard X-ray source spending most of its time in an out-of-outburst state with an average flux of $\sim \SI{1.4E-11}{\erg\per\second\per\centi\meter\squared}$.
The source occasionally exhibits X-ray flares lasting from a few hours to a few days, with a dynamic range of $20-200$ above $\SI{20}{\kilo\electronvolt}$ \citep{Sguera:2011}.
Since \isrc{} is not reported in the second INTEGRAL/IBIS catalogue \citep{Bird_Integral_IBIS_catalog2_2006}, despite an effective on-source exposure of $\SI{1.5}{\mega\second}$, the dynamic range above $\SI{20}{\kilo\electronvolt}$ can be increased to $> 900$.
\xmm{} observations further increased the dynamic range to $> 2500$ in the $0.5-10$ keV band \citep{Bozzo:2012}.
These X-ray properties supported the classification of \isrc{} as an intermediate SFXT, a subclass characterised by dynamic ranges and average luminosities between that of classical SFXTs and persistent supergiant HMXBs.
\emph{Chandra} observations improved the source localisation \citep{Tomsick2009:cxc17354}, enabling the spectroscopic identification of the optical/infrared counterpart\footnote{2MASS~J17352760$-$3255544, the companion donor star.} as a O9 supergiant \citep{Coleiro:2013}.
This firmly confirmed the classification of \isrc{} as an SFXT.
The Gaia mission provided accurate and reliable distance estimates of this counterpart \citep{Gaia_Mission_2016}.
Using Gaia EDR3 data \citep{Gaia_EDR3_2021} and the distance estimates from \citet{BailerJones_distances_2021}, the distance of \isrc{} is estimated at $4.1^{+1.7}_{-0.6}\si{\kilo\parsec}$.
\swift/BAT \citep{Dai:2011} and INTEGRAL \citep{Sguera:2011} observations revealed an orbital period of $8.448\pm0.002$ days.
The system likely has a low eccentricity ($\sim 0.1-0.2$).
Studies with \xmm{} \citep{Bozzo_2017} and INTEGRAL \citep{Goossens_2018} suggest that the flares of \isrc{} result from the accretion of dense clumps, rather than transitions in accretion mode.
Since these flares are of intermediate luminosity and are associated with modest variations in the absorption column density, the system might be able to easily overcome physical mechanisms that inhibit accretion.

\isrc{} was also proposed as the best candidate counterpart to the \gray transient AGL~J1734$-$3310 \citep{AGL2009Tel:Bulgarelli, Sguera:2011, sguera_IGRJ17354-3255_2013}.
This source was discovered at energies $>\SI{100}{\mega\electronvolt}$ by the AGILE \citep[Astrorivelatore Gamma ad Immagini LEggero, ][]{2009A&A...502..995T} mission on April 14, 2009, during a one-day flare.
The proposed association between \isrc{} and AGL~J1734$-$3310 is merely based  on circumstantial evidence, such as positional association and flaring activity on similarly short timescales.

In this work, we conduct a comprehensive search for \gray emission from the \isrc{} region using the entire AGILE/GRID data archive, from 2007 to 2024, in the $\SI{100}{\mega\electronvolt}-\SI{10}{\giga\electronvolt}$ energy range.
We improve the localisation of the HE transient, now renamed \asrc, we discuss its HE emission, characterised by multiple \gray flares, and we analyse INTEGRAL and \swift{} data from the same region.
In Sect.~\ref{sect:region}, we discuss the characterisation of the \isrc{} region.
In Sect.~\ref{sect:agile_observations} we describe the AGILE observations and report the data analysis methods and results in Sect.~\ref{sect:agile_results}.
In Sect.s~\ref{sect:integral} and \ref{sect:swift}, we report the data analysis and results of INTEGRAL and \swift{} observations, respectively.
In Sect.~\ref{sect:discussion}, we discuss our results, while our conclusions are reported in Sect.~\ref{sect:conclusions}.

\section{The \texorpdfstring{\isrc{}}{IGR~J17354-3255} region}
\label{sect:region}
\begin{figure}[t!]
 \centering
 \includegraphics[width=\columnwidth,keepaspectratio]{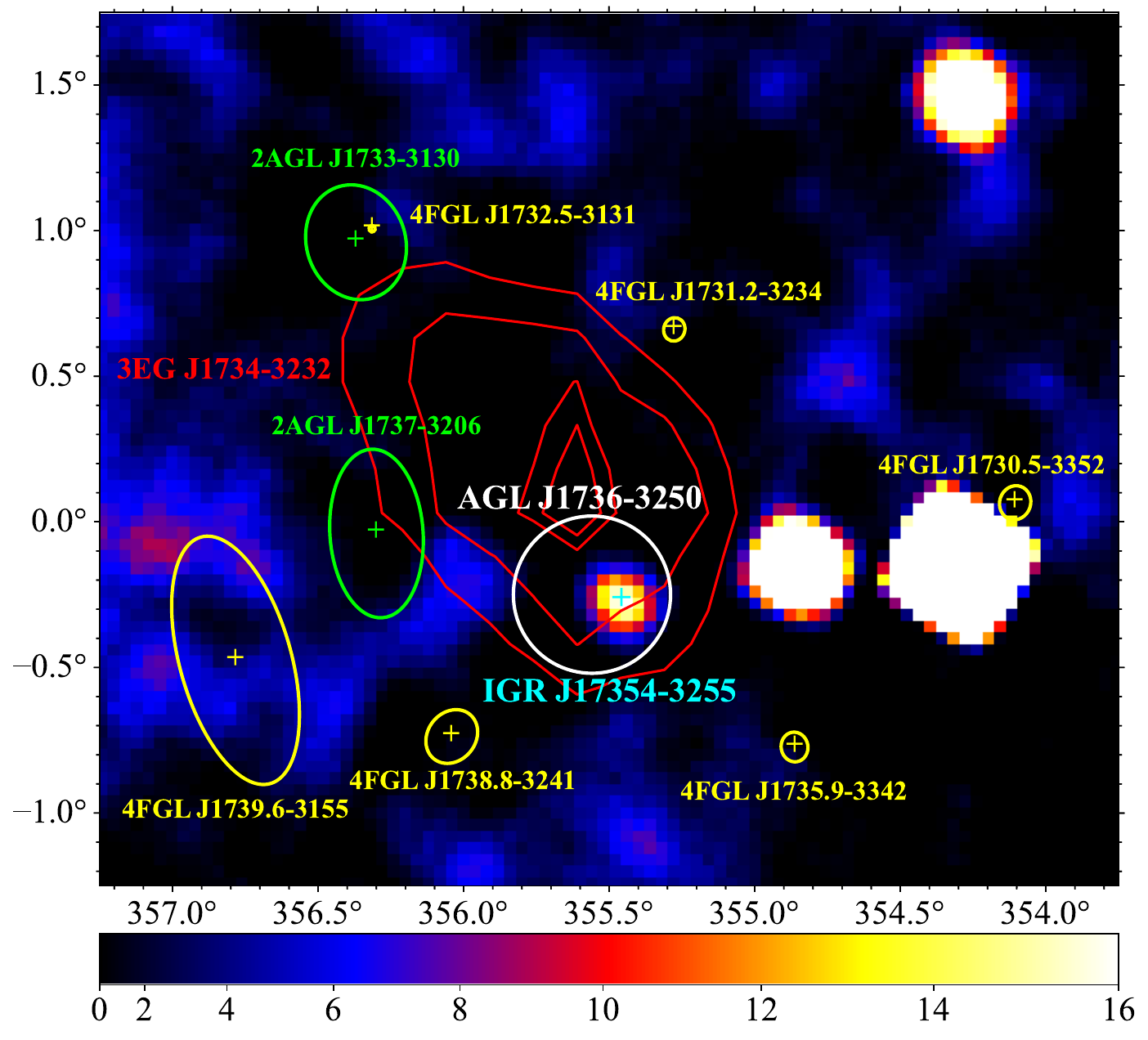}
 \caption{INTEGRAL/IBIS mosaic significance map ($18-60$ keV) of the sky region around \isrc{} in Galactic coordinates.
 The refined positional uncertainty of \asrc{} is shown by the white circle ($95\%$ confidence level), with 2AGL sources in green ellipses and \textit{Fermi}/LAT sources in yellow ellipses ($95\%$ confidence level).
 \isrc{}, marked in cyan, is the only hard X-ray source detected ($18\sigma$, $\SI{10}{\mega\second}$ effective on-source exposure time) unambiguously located inside the error circle of \asrc.
 The other two bright INTEGRAL/IBIS sources close to \isrc{} are the Low Mass X-ray binaries (LMXBs) GX~354$-$0 and 4U~1730$-$335.
 The red contours (from $50\%$ to $99\%$) refer to the EGRET source 3EG~J1734$-$3232.
 }
 \label{fig:skymap1}
\end{figure}

\isrc{} is located in a region of the sky densely populated with other hard X-ray sources.
However, it is the only hard X-ray object located within the error box of \asrc, both in the original localisation\footnote{$(\text{R.A.}, \text{Dec})=(\ang[angle-symbol-degree=\textsuperscript{h}, angle-symbol-minute=\textsuperscript{m}, angle-symbol-second=\textsuperscript{s}]{17;34;44},\ang{-33;10;38})$} \citep{AGL2009Tel:Bulgarelli} and in the updated coordinates inferred in Sect.~\ref{sect:AGILE_stacked_flare_analysis} (see Fig.~\ref{fig:skymap1}).

The region surrounding \isrc{} is also densely populated at $\si{\mega\electronvolt}$ and $\si{\giga\electronvolt}$ energies.
Within $\SI{1.5}{\deg}$, there is one EGRET source \citep[3EG~J1734$-$3232, ][]{1999ApJS..123...79H}, two \gray sources from the second AGILE catalogue \citep{Bulgarelli2019bt} and $11$ \gray sources from the Fourth \textit{Fermi}/LAT Catalogue \citep[4FGL-DR4, ][]{Fermi_4FGL_2020, Ballet_4FGL_DR4_2023}.
We show in Fig.~\ref{fig:skymap1} the main \gray sources surrounding \asrc, superimposed on the INTEGRAL/IBIS \citep{Ubertini_IBIS_2003} significance map of the region obtained in Sect.~\ref{sect:integral}.
None of the 4FGL-DR4 nor the 2AGL sources near \isrc{} are spatially associated with it or with \asrc.

The unidentified EGRET source 3EG~J1734$-$3232 is positionally consistent with both \asrc{} and \isrc{}.
It is listed in the third EGRET catalogue as a `confused' source, detected with significance of $\sqrt{TS} = 6.2$ and an average flux ($E > \SI{100}{\mega\electronvolt}$) of $(40.0\pm 6.7)\cdot 10^{-8}\si{\photon\per\second\per\centi\meter\squared}$ \citep{1999ApJS..123...79H}.
Its `confused' designation implies that the localisation may have significant uncertainties due to overlapping point spread functions (PSFs), suggesting that the source is likely a blend of several \gray sources.
This hypothesis is strongly supported by AGILE and \textit{Fermi} observations since, thanks to their excellent angular resolution, they have pinpointed the sources likely responsible for the emission of 3EG~J1734$-$3232; namely, \asrc{} and the 4FGL~J1732.5$-$3131 pulsar.
This is further supported by the elongated shape of 3EG~J1734$-$3232, which points towards the directions of the AGILE and \textit{Fermi} sources.
The hypothesis that the transient source \asrc{} contributes to the emission of the EGRET source is also supported by the measurements of the I index \citep{2000A&A...357..957Z} of 3EG~J1734$-$3232, which suggests that it is a likely \gray variable source \citep{2004A&A...421..983Z}.

\section{AGILE observations}
\label{sect:agile_observations}
AGILE is a \gray and X-ray astrophysics mission by the Italian Space Agency (ASI) that operated from April 23, 2007, to February 14, 2024 \citep{2008NIMPA.588...52T, 2009A&A...502..995T}.
Its instrument devoted to \gray imaging is the Gamma-Ray Imaging Detector (GRID), operating in the $\SI{30}{\mega\electronvolt}-\SI{50}{\giga\electronvolt}$ energy range.
For details about AGILE and its operations, we refer to \citet{Barbiellini2002jw, Prest2003280, Bulgarelli2010213, Cattaneo2011251,Feroci2007gk,2009NIMPA.598..470L,2006NIMPA.556..228P,Bulgarelli:2013gx, 2019RLSFN..30S.131V, Pittori:2019cp}.
AGILE data is available at the AGILE Data Center\footnote{\url{https://agile.ssdc.asi.it}}.
The AGILE spacecraft operated in `pointing mode' from the beginning of the mission to October 15, 2009, completing $101$ observation blocks (OBs).
The OBs\footnote{\url{https://agile.ssdc.asi.it/current_pointing.html}} usually consisted of predefined long exposures, drifting about $\SI{1}{\deg}$ per day with respect to the initial boresight direction to obey solar panels constraints.
In November 2009, the attitude control system was reconfigured, and scientific operations were performed in `spinning mode' until the end of the mission.
AGILE scanned $\approx 80\%$ sky daily (exposure of $\approx \SI{7E6}{\centi\meter\second}$) with an angular velocity of about $\SI{0.8}{\deg\per\second}$, performing $200$ passes per day on the same sky region.

A preliminary analysis of the AGILE data was performed using the public AGILE-LV$3$ web tool at SSDC \footnote{\url{https://www.ssdc.asi.it/mmia/index.php?mission=agilelv3mmia}}.
The results presented and discussed in this work are based on a refined analysis of the entire AGILE consolidated archive, publicly available from SSDC.
In Table~\ref{tab:agileobs0}, we report the AGILE observations of the \isrc{} region.
The dataset includes observations in both pointing (September 1, 2007-October 15, 2009) and spinning (Nov 1, 2009-December 31, 2023) modes. 
During the pointing mode period, AGILE observed the \isrc{} region repeatedly for a total of $\num{174}$ days.
For the spinning mode period, we analysed only those sub-periods that maximise the exposure to the source (daily exposure $>\SI{3.456E6}{\centi\meter\squared\second}$), resulting in the analysis of a fraction of the entire AGILE spinning period, amounting to $\num{1819}$ days.
Thus, we analysed a total of $\num{1993}$ days.

\begin{table}[ht!]
 \centering
 \caption{AGILE observations of the \isrc{} region.
 }
 \label{tab:agileobs0}
 \resizebox{\columnwidth}{!}{
 \begin{tabular}{ccccc}
  \hline\hline
  AGILE OB & $l$ & $b$ & Distance & Observation time \\
   & $[\si{\deg}]$ & $[\si{\deg}]$ & $[\si{\deg}]$ & [MJD] \\
  \hline
  1800  & 313.1 &  -3.5 & 42.4 & 54335.5 - 54339.5 \\
  1900  & 334.4 &  10.1 & 23.3 & 54339.5 - 54344.5 \\
  4400  &  19.3 & -15.4 & 28.0 & 54386.5 - 54395.5 \\
  5400  & 332.1 &   0.0 & 23.4 & 54526.5 - 54541.5 \\
  5450  & 360.0 &   0.6 &  4.6 & 54541.5 - 54555.5 \\
  6200  & 355.5 &   7.4 &  7.7 & 54719.5 - 54749.5 \\
  6800  & 349.9 &  13.4 & 14.8 & 54890.5 - 54921.5 \\
  7010  &  18.1 & -13.8 & 26.2 & 54928.5 - 54936.5 \\
  7800  & 307.3 &   0.5 & 48.1 & 55055.5 - 55074.5 \\
  7900  & 343.1 &  11.1 & 16.8 & 55074.5 - 55084.5 \\
  8200  &   3.1 &   5.3 &  9.5 & 55090.5 - 55104.5 \\
  8300  &  10.1 &  -6.5 & 15.9 & 55104.5 - 55119.5 \\
  \hline
  Spinning  &  - & -  & - & 55197.0 - 60309.0  \\
  \hline
 \end{tabular}
 }
  \tablefoot{The columns show: (1) AGILE observation block (OB) number for pointing-mode observations; (2-3) Galactic coordinates $(l,b)$ of the pointing centroids (in degrees); (4) distance of the pointing from \isrc{}; (5) observation time interval in MJD.
 We selected a total of $\num{1993}$ days of data, $174$ of which in pointing mode and $\num{1819}$ in spinning mode that maximise the exposure on the source.}
\end{table}

\section{AGILE/GRID data analysis and results}
\label{sect:agile_results}
We performed the analysis of AGILE/GRID \gray data using the FM3.119 on-ground background event filter, instrument response functions (IRFs) H0025 and the AGILE/GRID Science Tools (version BUILD25) publicly available at the ASI Science Data Center website.
We applied a calibrated filter for \gray events to account for South Atlantic Anomaly event cuts and $\SI{80}{\degree}$ Earth albedo filtering.
The GRID event direction was reconstructed using a Kalman filter technique.
To reduce particle background contamination, we selected only events flagged as confirmed \gray events (G class events).
We generated the AGILE counts, exposure, and Galactic background maps over a $\SI{30}{\deg}$ region centred on the position of \isrc, with a bin size of $\SI{0.25}{\degree} \times \SI{0.25}{\degree}$ for $E>\SI{100}{\mega\electronvolt}$.
We employed the standard AGILE binned multi-source likelihood analysis method to assess the statistical significance of the sources in the target region and evaluate their period-averaged flux and evolution \citep{Chen:2013er}.
The software iteratively optimised the position and spectrum of all the sources in the target region to search for both persistent and transient emission.
The region's emission model accounts for the Galactic diffuse \gray emission and the isotropic emission, that we modelled as fixed components with their average weekly level.
These settings represent the standard hypothesis for AGILE data analysis.
We used \texttt{agilepy v1.6.3}\footnote{\url{https://agilepy.readthedocs.io}} as the front-end software.
A complete description of the AGILE/GRID instrument, response characteristics, data analysis, and observation strategies is provided in \citet{Bulgarelli2019bt} and \citet{bulgarelli_agilepy_2022}.

The source model adopted for the analysis included a total of $22$ sources.
The first source represents \asrc, the hypothesised counterpart of \isrc, modelled as a point-like source with a power-law spectrum and a spectral index of $2.1$ (the standard AGILE power-law model).
The remaining $21$ are 2AGL sources detected within a $\SI{15}{\deg}$ radius of \isrc{}, for which we fixed the position and spectra.

We performed four types of analysis: a search for transient \gray flares on fixed timescales (Sect.~\ref{sect:1daytransientagile}), the stacked analysis of all detected flaring episodes (Sect.~\ref{sect:AGILE_stacked_flare_analysis}), a phase-folded analysis assuming the orbital period of \isrc{} for \asrc{} (Sect.~\ref{sect:AGILE_folded_analysis}), and a search for periodic emission of the HE emission (Sect.~\ref{sect:periodic}).

\subsection{Search for transient \gray emission}
\label{sect:1daytransientagile}
We evaluated the $0.1-10$~GeV flux of the AGILE/GRID source at the \isrc{} position and its evolution by computing the light curve for the timescale of one day, which is consistent with the characteristic outburst episodes of SFXTs in the hard X-ray band.
The spectral index was fixed at value $2.1$.
Our analysis identified $19$ bins with $\sqrt{TS} \ge 3.3$ (corresponding to a $3\sigma$ statistical significance; see Appendix~\ref{sect:appendix2}) representing positionally consistent \gray flaring episodes, reported in Table~\ref{tab:agileobs1}.
Eight of the flares were detected during `pointing mode' observations, which covered $174$ days, while $11$ during `spinning mode' observations, which spanned $\num{1819}$ days.
The lower detection rate in `spinning mode' can be attributed to its reduced coverage of the target region, which is available for only a fraction of the total observing time, limiting the ability of AGILE to monitor the variability of a source on hour-long timescales.
The flare with highest detection significance, labelled E05, was previously reported by \citet{AGL2009Tel:Bulgarelli}.

\begin{table*}
 \caption{Flares detected by AGILE/GRID from \asrc{} in the $\SI{100}{\mega\electronvolt}$ $-$ $\SI{10}{\giga\electronvolt}$ energy range in the one-day light curve, and simultaneous INTEGRAL observations available.
 }
 \label{tab:agileobs1}
 \resizebox{\textwidth}{!}{
 \begin{tabular}{ccccc|c|p{.16\textwidth}c}
  \hline\hline
  AGILE & & & & & & INTEGRAL & \\
  Flare & Observation date & $\sqrt{TS}$ & Photon Flux & Luminosity & Orbital & Observation date & Luminosity \\
  ID & [MJD] & & $[\SI{E-6}{\photon\per\second\per\centi\meter\squared}]$ & $[\SI{E37}{\erg\per\second}]$ & phase & [MJD] & $[\SI{E35}{\erg\per\second}]$\\
  \hline
  E01 & 54343.0 - 54344.0 & 3.5 & $1.75 \pm  0.63$ & $ 0.66 \pm 0.24$ & 0.77 & 54343.1 - 54344 (12 ks) &  $<$1.2 \\
  E02 & 54732.0 - 54733.0 & 3.4 & $1.51 \pm  0.57$ & $ 0.57 \pm 0.21$ & 0.82 & - & - \\
  E03 & 54746.0 - 54747.0 & 3.4 & $1.46 \pm  0.55$ & $ 0.55 \pm 0.21$ & 0.48 & - & - \\
  E04 & 54915.0 - 54916.0 & 4.2 & $2.19 \pm  0.66$ & $ 0.82 \pm 0.25$ & 0.48 & - & - \\
  E05 & 54935.0 - 54936.0 & 5.7 & $3.33 \pm  0.75$ & $ 1.25 \pm 0.28$ & 0.85 & - & - \\
  E06 & 55100.0 - 55101.0 & 3.3 & $1.21 \pm  0.49$ & $ 0.45 \pm 0.18$ & 0.38 & - & - \\
  E07 & 55109.0 - 55110.0 & 3.6 & $1.92 \pm  0.71$ & $ 0.72 \pm 0.27$ & 0.45 & - & - \\
  E08 & 55112.0 - 55113.0 & 3.6 & $1.84 \pm  0.65$ & $ 0.69 \pm 0.24$ & 0.80 & - & - \\
  \hline
  E09 & 55599.0 - 55600.0 & 3.7 & $3.07 \pm 1.30$ & $ 1.15 \pm 0.49$ & 0.45 & 55599.5 - 55600 (3 ks) &  $<$2.6 \\
  E10 & 56182.0 - 56183.0 & 3.3 & $2.36 \pm 1.07$ & $ 0.88 \pm 0.40$ & 0.47 & - & - \\
  E11 & 56199.0 - 56200.0 & 3.5 & $3.42 \pm 1.36$ & $ 1.28 \pm 0.51$ & 0.48 & 56199 - 56200 (9 ks) &  $<$1.5 \\
  E12 & 56351.0 - 56352.0 & 3.3 & $3.10 \pm 1.90$ & $ 1.16 \pm 0.71$ & 0.48 & 56351.6 - 56351.8 (3 ks)  &  $<$2.6 \\
  E13 & 57786.0 - 57787.0 & 4.3 & $5.55 \pm 1.89$ & $ 2.08 \pm 0.71$ & 0.35 & - & - \\
  E14 & 57823.0 - 57824.0 & 3.9 & $4.46 \pm 1.52$ & $ 1.67 \pm 0.57$ & 0.73 & 57823 - 57824 (5 ks) &  $<$2.2 \\
  E15 & 58199.0 - 58200.0 & 3.6 & $3.32 \pm 1.36$ & $ 1.24 \pm 0.51$ & 0.24 & - & - \\
  E16 & 58368.0 - 58369.0 & 3.3 & $3.02 \pm 1.34$ & $ 1.13 \pm 0.50$ & 0.24 & - & - \\
  E17 & 58744.0 - 58745.0 & 4.0 & $3.74 \pm 1.38$ & $ 1.40 \pm 0.52$ & 0.76 & - & - \\
  E18 & 59656.0 - 59657.0 & 3.3 & $2.95 \pm 1.31$ & $ 1.10 \pm 0.49$ & 0.72 & - & - \\
  E19 & 59806.0 - 59807.0 & 3.4 & $4.21 \pm 1.83$ & $ 1.57 \pm 0.69$ & 0.48 & - & - \\
\hline
 \end{tabular}
 }
 \tablefoot{
  When \isrc{} was within the INTEGRAL field of view ($\sim \SI{8}{\deg}$), we report the observations in the $18-60$ keV energy range simultaneous to AGILE.
 The columns show: (1) AGILE flare ID; (2) AGILE integration time in MJD; (3) the statistical significance expressed in $\sqrt{TS}$ of the source detection according to the maximum likelihood ratio test; (4) the average photon flux and its $1\sigma$ statistical error, evaluated in the $\SI{100}{\mega\electronvolt}$ $-$ $\SI{10}{\giga\electronvolt}$ range assuming a power-law spectrum with index $2.1$; (5) the corresponding luminosity assuming a distance of $\SI{4.1}{\kilo\parsec}$; (6) the orbital phase centred in the integration time, considering the zero phase ephemeris at MJD 52698.205 and that the orbital period of the binary system is $8.4474 \pm 0.0017$ days; (7) INTEGRAL/IBIS observations time coverage (MJD) of the \gray flare duration and on-source effective exposure; and (8) INTEGRAL 3$\sigma$ upper limit ($18-60$ keV) in $\SI{E35}{\erg\per\second}$ .
 The horizontal line separates the `pointing mode' and the `spinning mode' periods.}
\end{table*}

To evaluate the statistical significance of the AGILE detections, we assessed the post-trial probability of multiple independent flares occurring from the same sky region within the analysis trials (i.e. the repeated flare occurrence).
The post-trial probability of detecting $19$ or more detections with significance $\ge 3\sigma$ out of $1993$ trials under the null hypothesis is $P = \num{7.448E-10}$, corresponding to $\sim 6.04\sigma$ Gaussian standard deviations.
Appendix~\ref{sect:appendix2} provide details on the pre-trial and post-trial significance evaluation, along with the $TS$ distribution of the \isrc{} region with the likelihood test performed.

In Table~\ref{tab:agileobs1}, we also report the average orbital phase of each \gray flare, assuming a zero phase ephemeris at MJD 52698.205 and an orbital period of $8.4474 \pm 0.0017$ days, consistent with \isrc{} \citep{Dai:2011}.
In Fig.~\ref{fig:folded}a we show the distribution of the detected \gray flares as a function of the binary system's orbital phase.
Approximately half of the \gray flares occurs near the apastron passages.
Notably, $8$ of the $19$ flares were detected during the orbital phase interval $[0.4375-0.500]$.
The probability of this clustering occurring by chance is $\num{8.65E-6}$, corresponding to $\sim 4.3\sigma$ Gaussian standard deviations.
These results suggest that \asrc{} emits a significant fraction of its \gray flares near the apastron of \isrc{}, while the other, smaller clustering of flares visible in Fig.~\ref{fig:folded}a occurs around phase $0.75$; the probability of detecting at least $7$ flares within the phase interval $[0.72-0.85]$ is $\num{7.56e-3}$ (corresponding to $\sim 2.5\sigma$).
This clustering, which includes the most significant flare (E05), may be a chance occurrence.

Since the variability of SFXT can occur on timescales of only a few hours, we investigated the intra-day behaviour of the flaring activity of \asrc{}.
For the $19$ days with flaring days listed in Table~\ref{tab:agileobs1}, we computed light curves using predefined binning of $\SI{2}{\hour}$, $\SI{4}{\hour}$, and $\SI{6}{\hour}$.
For each daily flaring episode, we selected the bins with $\sqrt{TS}\geq 3.3$ and identified the one with the highest significance across the three light curves.
We derived the flux and and the photon excess associated with the short flare using a maximum likelihood analysis.
The most significant intra-day flare for each episode is reported in Table~\ref{tab:agileobs3}.
In most cases, we found that $\approx 25-50\%$ of the photons detected in the day-long bursts were emitted within a shorter timescale.
This dynamic behaviour, where emission is concentrated in fast, hour-long flares with a low duty cycle, is typical of SFXTs in hard X-ray band.
Our analysis indicates that this behaviour is also present in the \gray band.

\begin{table*}
 \caption{Most significant intra-day flare from \asrc{} in the $\SI{100}{\mega\electronvolt}$ $-$ $\SI{10}{\giga\electronvolt}$ energy range during each of the $1-$day flaring episodes reported in Table~\ref{tab:agileobs1}.
 }
 \label{tab:agileobs3}
 \resizebox{\textwidth}{!}{
 \begin{tabular}{cc|cccccc}
  \hline\hline
  1-day flare & & Most significant intra-day flare & & & & & \\
  Flare & Counts & Integration Time & Duration & $\sqrt{TS}$ & Counts & Photon Flux & Luminosity \\
  ID & $\si{\photon}$ & MJD & $\si{\hour}$ & & $\si{\photon}$ & $[\SI{E-6}{\photon\per\second\per\centi\meter\squared}]$ & $[\SI{E37}{\erg\per\second}]$ \\
  \hline
  E01 & $33 \pm 12$ & 54343.000 - 54343.083 & 2 & 3.5 & $11 \pm 5  $ & $10.09 \pm 4.36$ & $3.78 \pm 1.63$ \\
  E02 & $31 \pm 12$ & 54732.250 - 54732.500 & 6 & 3.1 & $19 \pm 8  $ & $ 2.96 \pm 1.25$ & $1.11 \pm 0.47$ \\
  E03 & $28 \pm 11$ & 54746.416 - 54746.500 & 2 & 3.3 & $11 \pm 5  $ & $ 6.46 \pm 3.03$ & $2.42 \pm 1.13 $ \\
  E04 & $44 \pm 13$ & 54915.333 - 54915.416 & 2 & 3.5 & $ 8 \pm 4  $ & $ 7.55 \pm 3.71$ & $2.83 \pm 1.39 $ \\ 
  E05 & $62 \pm 14$ & 54935.666 - 54936.750 & 2 & 4.1 & $15 \pm 6  $ & $ 6.89 \pm 2.66$ & $2.58 \pm 1.00 $ \\
  E06 & $26 \pm 10$ & - &  &  &  &  &  \\
  E07 & $27 \pm 10$ & 55109.000 - 55109.166 & 4 & 4.3 & $19 \pm 7  $ & $ 5.04 \pm 1.79$ & $1.89 \pm 0.67 $  \\
  E08 & $37 \pm 13$ & 55112.583 - 55112.666 & 2 & 3.6 & $12 \pm 5  $ & $ 6.64 \pm 2.76$ & $2.48 \pm 1.03 $ \\
  \hline
  E09 & $14 \pm 6$  & 55599.000 - 55599.083 & 2 & 3.9 & $ 7 \pm 4$ & $13.14 \pm 6.80$ & $4.92 \pm 2.54 $\\
  E10 & $15 \pm 7$  & 56182.000 - 56183.250 & 6 & 3.5 & $ 7 \pm 4$ & $ 4.57 \pm 2.35$ & $1.71 \pm 0.88 $ \\
  E11 & $18 \pm 7$  & - &  &  &  & & \\
  E12 & $12 \pm 6$  & 56351.500 - 56351.666 & 4 & 3.3 & $ 7 \pm 4$ & $ 7.37 \pm 3.87$ & $2.76 \pm 1.45 $  \\
  E13 & $21 \pm 7$  & 57786.000 - 57786.250 & 6 & 3.2 & $ 8 \pm 4  $ & $ 7.57 \pm 3.84$ & $2.83 \pm 1.44 $ \\
  E14 & $27 \pm 9$  & - &  &  &  & & \\
  E15 & $19 \pm 8$  & 58199.333 - 58199.500 & 4 & 3.9 & $11 \pm 5  $ & $10.69 \pm 4.69$ & $4.00 \pm 1.76 $\\
  E16 & $16 \pm 7$  & 58368.000 - 58368.166 & 4 & 3.6 & $ 8 \pm 4  $ & $ 9.98 \pm 4.83$ & $3.73 \pm 1.81 $ \\
  E17 & $21 \pm 8$  & 58744.000 - 58744.250 & 6 & 4.0 & $12 \pm 5  $ & $ 8.45 \pm 3.52$ & $3.16 \pm 1.32 $  \\
  E18 & $16 \pm 7$  & -  &  &  &  & & \\
  E19 & $14 \pm 6$  & 59806.250 - 59806.500 & 6 & 3.3 & $ 7 \pm 4$ & $ 6.18 \pm 3.30$ & $2.31 \pm 1.23 $  \\
  \hline
 \end{tabular}
 }
 \tablefoot{Column (1) provides the flare ID, (2) the excess counts associated with \asrc{} in Table~\ref{tab:agileobs1}.
 The other columns are relative to the intra-day flare and provide: (3) integration range in MJD; (4) integration time in hours; (5) statistical significance expressed in $\sqrt{TS}$ of the detection according to the maximum likelihood ratio test; (6) number of excess counts and $1\sigma$ statistical error associated with \asrc{}; (7) the average photon flux and its $1\sigma$ statistical error, evaluated assuming a power-law spectrum with index $2.1$; and (8) the corresponding luminosity assuming a distance of $\SI{4.1}{\kilo\parsec}$.
 The horizontal line separates the `pointing' and the `spinning' mode periods.}
\end{table*}

\subsection{AGILE-Fermi comparison}
\label{sect:AGILE_Fermi}

No \gray{} source consistent with the position of \isrc{} is reported in the \textit{Fermi} catalogue \citep{Fermi_4FGL_2020} or in \textit{Fermi}-LAT transient activity notices.
This is not unprecedented, as other transient sources detected by AGILE have remained undetected by \textit{Fermi} despite its larger average effective area; see for instance \citet{alexander_BHbinary_2015} and \citet{MunarAdrover:2016:MWC656}.
These previous works have highlighted the conditions under which short flares may escape detection by \textit{Fermi}-LAT.
In practice, the detectability of transient emission depends not only on the effective area, but also on the source offset relative to the telescope pointing direction and the observing mode.
Altogether, these properties determine the effective exposure time.

\begin{figure}[t!]
 \centering
 \includegraphics[width=\columnwidth,keepaspectratio]{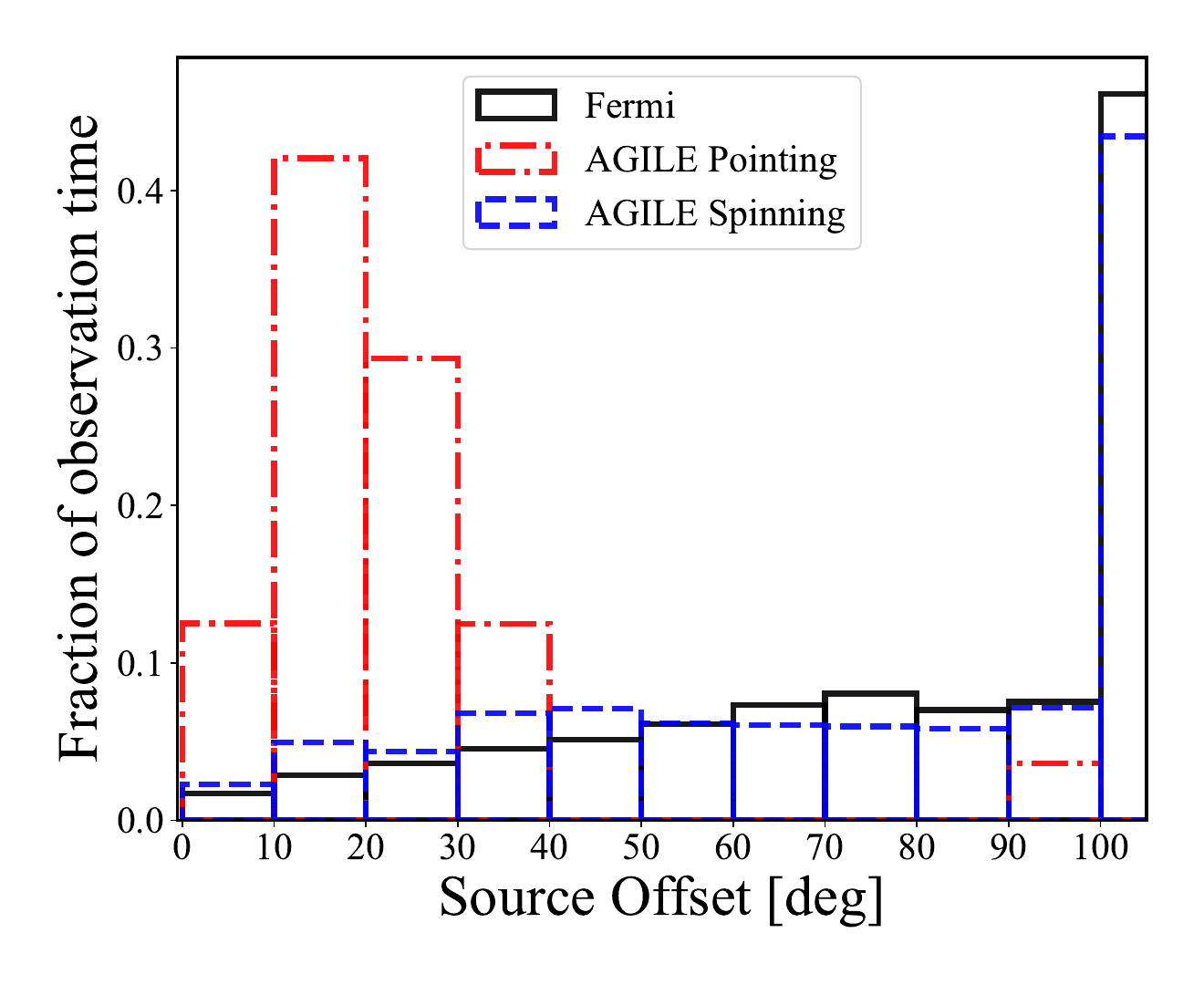}
 \caption{Distribution of \asrc{} offset angles during the flaring days listed in Table~\ref{tab:agileobs1}, for AGILE pointing (dash–dotted red line), spinning (dashed blue line), and \textit{Fermi} (solid black line) observations.
 The distribution is expressed as fraction of the observation time.
 }
 \label{fig:fermi_agile_offset}
\end{figure}

\begin{figure}[t!]
 \centering
 \includegraphics[width=\columnwidth,keepaspectratio]{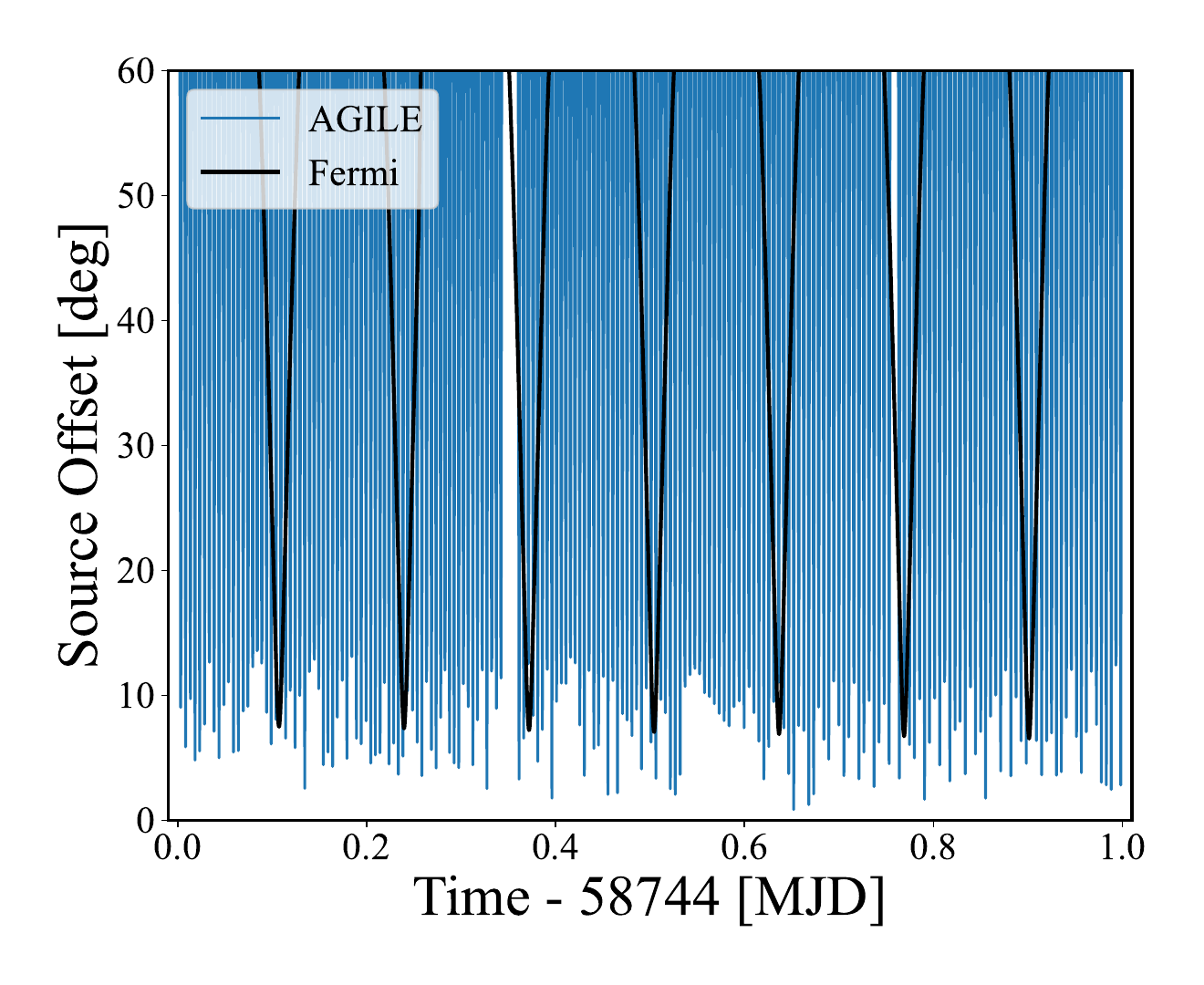}
 \caption{Offset angle of \asrc{} for \textit{Fermi} (black line) and AGILE (blue line) during flare E17, starting at MJD~$58744$.
 }
 \label{fig:fermi_agile_visibility}
\end{figure}

We computed the off-axis angles of \asrc{} with respect to the pointing directions of \textit{Fermi}-LAT and AGILE-GRID during the $19$ days, listed in Table~\ref{tab:agileobs1}, and during the most significant intra-day flares in Table~\ref{tab:agileobs3}.
Fig.~\ref{fig:fermi_agile_offset} compares the offset distributions for \textit{Fermi}, AGILE pointing, and AGILE spinning observations.
On average, \textit{Fermi}/LAT observed the source with a duty cycle (fraction of time the source was observed with an offset $<\SI{50}{\deg}$) $d_{50} \approx 18\%$ and a mean offset within that interval of $\SI{32.3}{\deg}$; similar values hold for the intervals of the most significant intra-day flares.
For AGILE, the results differ significantly between pointing and spinning observing modes.
Pointing observations achieve $d_{50} = 100\%$ with average offsets around $\SI{18.4}{\deg}$.
In contrast, spinning mode observations yield $d_{50} \approx 25\%$ and mean offsets of $\SI{30.6}{\deg}$, similar to \textit{Fermi}.
As an illustration, Fig.~\ref{fig:fermi_agile_visibility} shows the offset of \asrc{} during E17, when both satellites achieved comparable duty cycles and mean offsets.

The main difference between AGILE and \textit{Fermi} is thus the presence of AGILE's pointed observations targeting the source after the initial 2009 flare detection, which provide continuous visibility required by our analysis.
A crucial aspect of our methodology described in Setion~\ref{sect:1daytransientagile} is that none of the individual AGILE flaring episodes is highly significant on its own, but the evidence of a detection emerges only cumulatively when these episodes are coherently stacked under the repeated-flare hypothesis.
This approach differs fundamentally from typical time-averaged or blind transient searches, where long integration periods, including many non-flaring intervals, can dilute short-duration signals below detectability.
In this cumulative framework, the AGILE pointed observations are the main contributors to the final detection.
When only pointed observations are included ($8$ flaring episodes over $174$ days), the total significance reaches $\sim 6\sigma$.
Using only spinning-mode episodes ($11$ over $1819$ days) yields a significantly lower cumulative significance of $\sim 3.8\sigma$.
Thus, it is AGILE's pointed strategy that allowed us to capture a set of short, weak flares that remain statistically marginal unless combined coherently.

Although these considerations did not enable us to confirm or exclude the presence of a flare in \textit{Fermi} data, they do demonstrate that suboptimal observing conditions could have prevented \textit{Fermi} from detecting the \asrc{} flares.
Additional differences might also arise from systematic effects in event classification, model selection, and background estimation, which are particularly critical for sources located on the Galactic plane such as \asrc{}.

A preliminary analysis of the photometric \textit{Fermi} light curves was conducted for the flare epochs and in the direction of \asrc{}; however, it did not yield conclusive results.
The light-curve fluxes do not show strong evidence of flaring activity, as expected due to contamination from at least three nearby sources (see Fig.~\ref{fig:skymap1}), large PSF, and limited photon statistics.
On the other hand, the photometric light-curve fluxes are consistent with the AGILE results.
This supports the need for a dedicated and thorough \textit{Fermi} analysis.

\subsection{Stacked analysis of flaring episodes}
\label{sect:AGILE_stacked_flare_analysis}
We performed a stacked analysis of the flares in Table~\ref{tab:agileobs1}, yielding a significance of $11\sigma$ for the stacked dataset.
The transient source is centred at Galactic coordinates $(l,b)=(355.56, -0.25)\,\si{\deg}$, corresponding to $(\text{R.A.}, \text{Dec})=(\ang[angle-symbol-degree=\textsuperscript{h}, angle-symbol-minute=\textsuperscript{m}, angle-symbol-second=\textsuperscript{s}]{17;35;39},\ang{-32;49;53})$.
The position has a $95\%$ error circle of $\SI{0.20}{\deg}$ (statistical) $\pm \SI{0.07}{\deg}$ (systematic).
Additionally, the error ellipse has axes of $(0.26, 0.14)\,\si{\deg} \pm \SI{0.07}{\deg}$, with a rotation angle of $\SI{17.9}{\deg}$ clockwise.
Based on these findings, we designated the AGILE transient as \asrc.
The revised position of \asrc{} is closer to \isrc{} than the position previously determined.

Thanks to the increased photon statistics, the spectral index of \asrc{} was left free to vary, yielding a best-fit value of $2.10 \pm 0.11$.
The average $0.1-10 \, \si{\giga\electronvolt}$ flux is $(1.89\pm 0.27)\cdot 10^{-6}\si{\photon\per\second\per\centi\meter\squared}$, corresponding to $(3.52\pm 0.50)\cdot 10^{-9}\si{\erg\per\second\per\centi\meter\squared}$.
Assuming a source distance of $\SI{4.1}{\kilo\parsec}$, the average \gray flare luminosity is $L_\gamma \simeq (7.1\pm1.0)\cdot 10^{36}\,\si{\erg\per\second}$.

In Fig.~\ref{fig:skymap1}, we show the INTEGRAL/IBIS mosaic significance map ($18-60$~keV) of the sky region surrounding \isrc, as obtained from the INTEGRAL/IBIS catalogue dataset \citep{2016ApJS..223...15B}.
The refined positional uncertainty of \asrc{} is overlaid.
\isrc{} is the only hard X-ray source detected by INTEGRAL within the AGILE error circle.

\begin{figure}[ht!]
\centering
\includegraphics[width=\columnwidth,keepaspectratio]{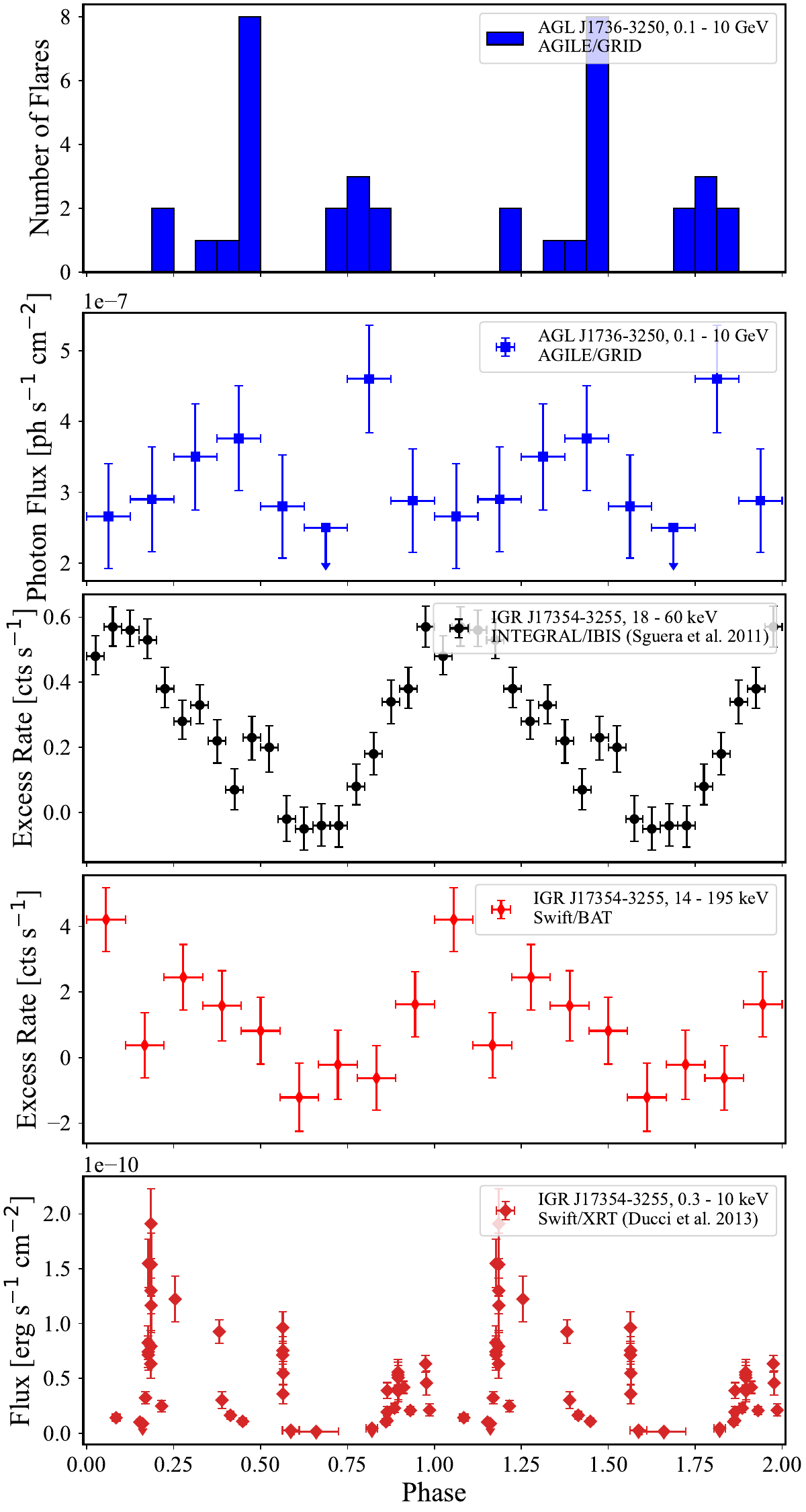}
\caption{Phase-folded light curves of \asrc{} and \isrc{}.
The zero phase ephemeris is MJD 52698.205 and the orbital period is $(8.4474\pm 0.0017)\si{\day}$. 
Phase $0$ is at periastron passage. 
From top to bottom: 
(a) Histogram of the number of flares detected by AGILE as reported in Table \ref{tab:agileobs1}; 
(b) AGILE/GRID $0.1-10$~GeV phase-folded light curve ($1\sigma$ errors and upper limits), that results variable at $99\%$ confidence level (see text);   
(c) INTEGRAL/IBIS $18-60$~keV phase-folded light curve of \isrc{} \citep{Sguera:2011}; 
(d) \swift/BAT $14-195$~keV phase-folded light curve of \isrc; 
(e) \swift/XRT $0.3-10$~keV phase-folded light curve of \isrc{} \citep{Ducci:2013}. 
\label{fig:folded}}
\end{figure}

\subsection{Phase-folded analysis}
\label{sect:AGILE_folded_analysis}
We searched for periodicity in the integrated data ($1993$ days with good exposure; see Sect.~\ref{sect:agile_observations}) folding the light curve of \asrc{} assuming the orbital period of \isrc{}.
In Fig.~\ref{fig:folded}b we present the eight-bin \gray phase-folded light curve of \asrc{} in the $\SI{100}{\mega\electronvolt} - \SI{10}{\giga\electronvolt}$ energy range.
The light curve shows a higher than average flux during the $[0.750-0.875]$ orbital phase.
This phase coincides with a clustering of flares, which is clearly visible in Fig.~\ref{fig:folded}a.
The $0.1-10 \, \si{\giga\electronvolt}$ flux obtained stacking the eight phase bins is $(3.08\pm 0.26)\cdot 10^{-7}\si{\photon\per\second\per\centi\meter\squared}$.

To evaluate variability, we computed the variability index VI \citep{Bulgarelli2019bt} comparing the AGILE/GRID phase-folded light curve to a constant model.
The VI index indicates \asrc{} is variable at $99\%$ confidence level.
However, this analysis does not distinguish whether the variability arises from periodic orbital modulation of the signal or non-periodic flaring activity.
These observations suggest that the \gray variability of \asrc{} is not driven by orbital modulation, but it instead due to flaring activity.

\subsection{Search for periodic \texorpdfstring{\gray}{gamma-ray} emission}
\label{sect:periodic}
We performed a timing analysis of \asrc{} data to assess whether the variability detected in Sect.~\ref{sect:AGILE_folded_analysis} could be attributed to phase modulation with a period of $8.4474$ days, corresponding to the orbital period of \isrc{}.
We created light curves using the aperture photometry technique on $\SI{1}{\hour}$, $\SI{2}{\hour}$, $\SI{4}{\hour}$, and $\SI{1}{\day}$ timescales, extracting photons from a circular region centred on X-ray coordinates of \isrc, with radii of $\SI{2}{\deg}$ and $\SI{3}{\deg}$ in the $0.1-10 \, \si{\giga\electronvolt}$ energy range.
The contribution of data points to the power spectrum was weighted by their relative exposures.

To search for periodicity, we applied the Lomb-Scargle (LS) periodogram, a widely used method for detecting sinusoidal periodic components in unevenly sampled time series \citep{Lomb:1976bo, Scargle:1982eu, VanderPlas:2018dw}.
The LS algorithm, implemented via the \texttt{AstroPy} package \citep{astropy:2013, astropy:2018, astropy:2022}, estimates the Fourier power spectrum as a function of the oscillation period or frequency.
No significant periodic components were detected in the data.
These results suggest that the \gray variability of \asrc{} is not driven by the orbital modulation of \isrc{} and it is instead likely due to non-periodic flares.

\section{INTEGRAL data analysis and results}
\label{sect:integral}
The INTEGRAL mission has monitored the \isrc{} region since its discovery in 2006, accumulating $\gtrsim \SI{10}{\mega\second}$ of exposure time on the source, listed in several INTEGRAL IBIS/ISGRI catalogues \citep{2007ApJS..170..175B, 2010ApJS..186....1B, 2016ApJS..223...15B, 2022MNRAS.510.4796K}.
In Fig.~\ref{fig:skymap1} we show the INTEGRAL IBIS/ISGRI mosaic significance map ($18-60$~keV) of the sky region around \isrc.
In Fig.~\ref{fig:folded}c we show the INTEGRAL IBIS/ISGRI folded light curve ($18-60$~keV) of \isrc{} published by \citet{Sguera:2011}, which reveals a smooth orbital modulation of the excess rate.
The emission peaks during periastron passage and drops to excess rate values consistent with zero near apastron.
The typical hard X-ray outbursts detected by INTEGRAL, with average X-ray luminosities of $L_X \sim \SI{e35}{\erg\per\second}$, are unlikely to produce the observed smooth orbital emission profile over long-baseline observations, as shown, for example, in Fig.~\ref{fig:folded}c.
Instead, this behaviour is attributed to low-intensity hard X-ray emission, typically below the sensitivity of INTEGRAL, which becomes detectable only when long-exposure data are folded.
The X-ray emission may thus represent a superposition of low-intensity activity with luminosities of $\sim 10^{33-34} \, \si{\erg\per\second}$ \citep{Sguera:2011}. 

We analysed the INTEGRAL IBIS/ISGRI archive to search for hard X-ray flaring activity from \isrc{} in the $18-60$~keV energy band simultaneous to the \gray flares detected by AGILE, using the latest ISDC offline scientific analysis software (version 11.2).
As shown in Table~\ref{tab:agileobs1}, \isrc{} was within the INTEGRAL field of view (FoV) by chance during $5$ of the $19$ \gray flares detected by AGILE, although with low effective on-source exposure times (ranging from $3-12$~ks.
We did not find any significant detections (i.e. $>5\sigma$) in the $18-60$~keV energy band for any of the flares.
In Table~\ref{tab:agileobs1}, we report the inferred $3\sigma$ upper limits on the hard X-ray luminosities, which are below values of the order of $L_X \lesssim \SI{2E35}{\erg\per\second}$.
We computed IBIS/ISGRI light curves with $\SI{2}{\kilo\second}$ bins in the $18-60$~keV range.
We note that INTEGRAL observations did not cover the most significant intra-day flares detected by AGILE in Table~\ref{tab:agileobs3}, and no $\SI{2}{\kilo\second}$-long flares were detected in the other bins of the IBIS/ISGRI 2ks light curve.

\section{Swift data analysis and results}
\label{sect:swift}
\swift/BAT identified a counterpart to \isrc{} \citep{Barthelmy2005:BAT}, catalogued as Swift~J1735.6$-$3255, in the 58-month hard X-ray survey \citep{Baumgartner2010:BAT58mos} and in the 54-month Palermo \swift/BAT hard X-ray catalogue \citep{Cusumano2010:batsur_III}.

For this study, we analysed the \swift/BAT data from the 70-Month Hard X-ray Survey \citep{Baumgartner2011:BAT70mos}, covering the MJD range 53355$-$55469 (December 16,2004 to September 30, 2010).
We produced a folded light curve of the source in the $14-195$~keV band, assuming the orbital period $P=\SI{8.4474}{\day} = \SI{729855.375}{\second}$ of \isrc{} and its periastron passage at MJD $52698.205$ as a zero phase ephemeris.
We show the folded light curve in Fig.~\ref{fig:folded}d, where we can see it exhibits emission modulation peaking at the periastron passage.

AGILE observations reported by \citet{AGL2009Tel:Bulgarelli} triggered several \swift{} pointed observations of the \isrc{} region.
These include observations on April 17, 2009 \citep{SWIFT2009Tel:Vercellone, Ducci:2013}.
Furthermore, a \swift/XRT and \swift/UVOT monitoring campaign comprising $25$ observations was performed in $11$ days between July 18 and July 28, 2012 \citep{Ducci:2013}.
No AGILE flare detected in Table~\ref{tab:agileobs1} was covered by these observations.
In Fig.~\ref{fig:folded}e we show the \swift/XRT folded light curve in the $0.3-10$~keV energy range published by \citet{Ducci:2013}, which exhibits the same orbital modulation observed by \swift/BAT \citep{Dai:2011}.

\begin{figure*}
 \sidecaption
 \includegraphics[width=12 cm,keepaspectratio]{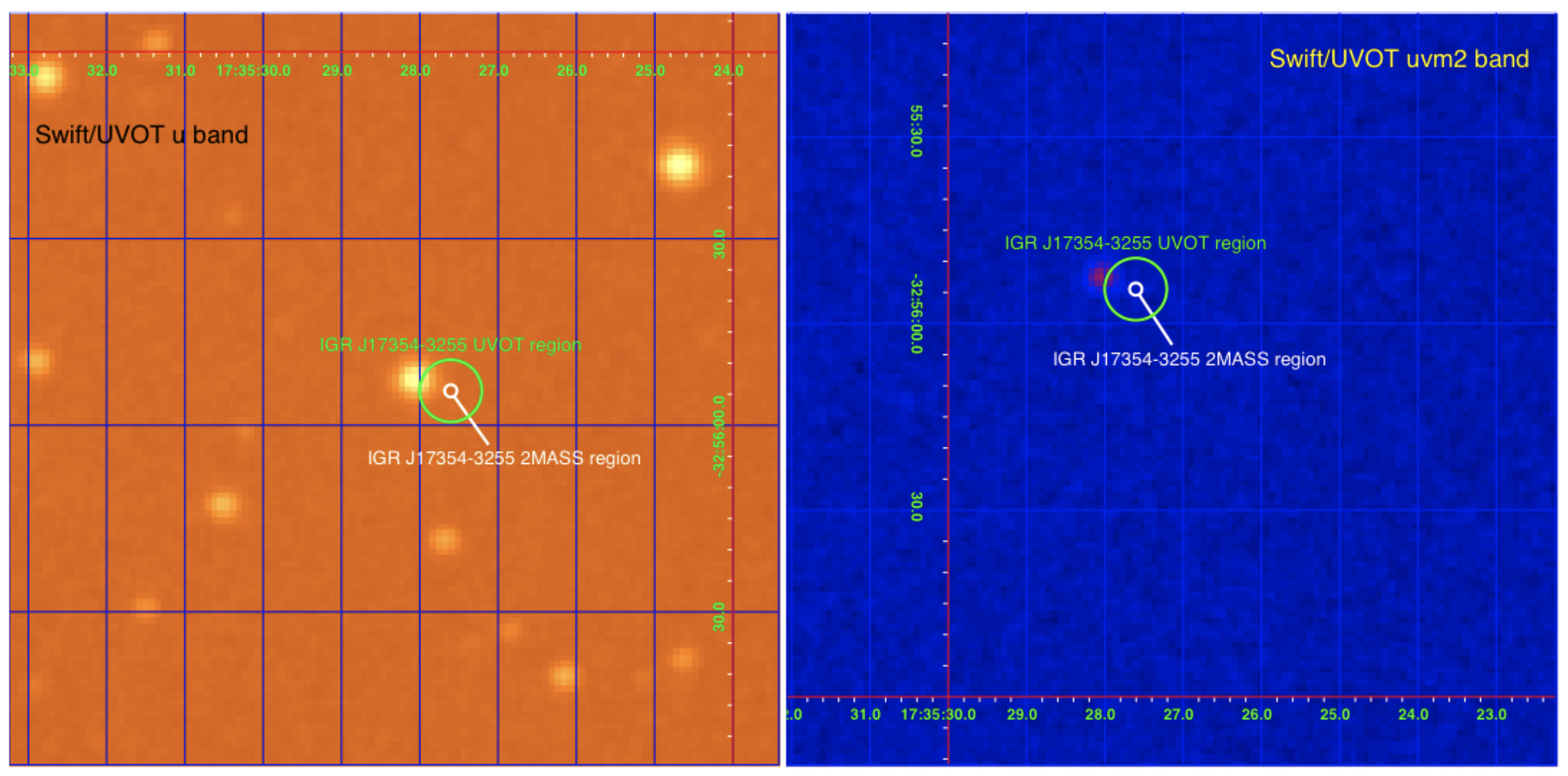}
 \caption{\swift/UVOT images in band U (left panel) and UVM2 (right panel) of \isrc.
 The green circles mark the $5\arcsec$ radius for the \swift/UVOT pipeline extraction region; the white circles mark the position of the optical/infrared counterpart of the source, the 2MASS~J17352760$-$3255544 star.
 The bright star inside the green circles is a contaminating source. 
\label{fig:swiftuvot}}
\end{figure*}

We also analysed the \swift/UVOT \citep{Roming2005:UVOT} data obtained simultaneously with the XRT observations using the {\sc uvotimsum} and {\sc uvotsource} tasks in {\sc FTOOLS}\footnote{\url{https://heasarc.gsfc.nasa.gov/ftools}} \citep{NASA_FTOOLS_2014}.
The {\sc uvotsource} task calculates the source magnitude through aperture photometry.
Magnitudes are provided in the Vega photometric system \citep{Poole2008:UVOT} and are not corrected for Galactic extinction.
We extracted the source counts from a circular region centred on \isrc{} with a radius of $5\arcsec$.
However, a nearby contaminating source within this radius precluded uncontaminated measurements, even when the extraction radius was reduced to $1\arcsec$.
We evaluated the background from source-free circular regions in the surroundings of the source.
Due to contamination, we computed $3\sigma$ upper limits for the source magnitude in the U and UVM2 bands.
During MJD $54536.2492 \pm 0.2043$, the U band magnitude upper limit was U$>20.81\,\text{mag}$ (flux $\phi_\text{U} < \SI{1.2E-14}{\erg\per\second\per\centi\meter\squared}$) and during MJD $54938.1900 \pm 0.1420$, the UVM2 band upper limit was UVM2$>20.93\,\text{mag}$ (flux $\phi_\text{UVM2} < \SI{8.7E-15}{\erg\per\second\per\centi\meter\squared}$).
In Fig.~\ref{fig:swiftuvot}, we show the \swift/UVOT images of \isrc{} in the U (left panel) and UVM2 (right panel) bands.

\section{Discussion}
\label{sect:discussion}
We analysed the entire (2007-2024) AGILE consolidated archive to search for \gray emission from the region of \isrc{}.
We report our results below.

\subsection{Detection and multi-wavelength analysis of \texorpdfstring{\gray}{gamma-ray} flares from \texorpdfstring{\asrc{}}{AGL~J1736-3250}}
\label{sect:discussion:episodic_flares}
\isrc{} is classified as a SFXT exhibiting X-ray outbursts on timescales typically from a few hours to a few days; hence, we searched for \gray flares over similar durations.
We identified $19$ days (see Table~\ref{tab:agileobs1}) that were characterised by a statistical significance $\geq 3\sigma$, yielding a $6.04\sigma$ detection under the hypothesis of repeated flare occurrence (see Appendix~\ref{sect:appendix2}).
In most cases, $\approx 25-50\%$ of photons detected in day-long flares were emitted over shorter timescales, as illustrated in Table~\ref{tab:agileobs3}.
This indicates that \asrc{} is a \gray source with a low duty cycle, concentrating its emission in fast, hour-long flares.
This dynamic behaviour mirrors that of \isrc{} in hard X-rays as detected by INTEGRAL. 

The AGILE/GRID analysis (Sect.~\ref{sect:1daytransientagile}) detected eight of the $19$ flares during AGILE's `pointing mode' observations, which feature a $70-88\%$ duty cycle and an off-axis angle of $7-32$~deg.
This mode allowed near-continuous observations, interrupted only by spacecraft passages through the South Atlantic Anomaly, and likely explains the higher detection rate during this period.
In contrast, AGILE's `spinning mode' has a $27-34\%$ duty cycle and off-axis angle of $27-34$~deg, providing less exposure to the target region.
Detecting such flares requires optimal time coverage, achievable only through continuous observations.
The observations conditions might be the reason why the \asrc{} was not detected by \textit{Fermi}-LAT in any catalogue nor notice of transient observations.

A stacked analysis of the $19$ flares confirmed the detection of the \gray source, named \asrc{}, with a $11\sigma$ significance level.
The only known hard X-ray source ($>\SI{20}{\kilo\electronvolt}$) located inside the error circle of \asrc{} is \isrc{} (see Fig.~\ref{fig:skymap1}), which suggests a possible positive association between the two sources.
The average \gray luminosity during flares is $L_\gamma \simeq(7.1\pm1.0)\cdot 10^{36}\si{\erg\per\second}$ ($0.1-10$~GeV, assuming a source distance of $\SI{4.1}{\kilo\parsec}$), while typical \gray luminosity values are in the range $(0.4-2)\cdot\SI{E37}{\erg\per\second}$.

\isrc{} was within the INTEGRAL FoV by chance during $5$ of the $19$ \gray flares with low effective on-source exposure times (in the range $3-12$~ks).
No significant emission was detected in the $18-60$~keV energy band, with an inferred $3\sigma$ luminosity upper limits of $\sim\SI{2E35}{\erg\per\second}$.
Under the hypothesis that \gray flares and hard X-ray flares are produced by the same physical mechanism, this implies that $L_\gamma \gg L_X$ by about one or two orders of magnitude.
Notably, typical hard X-ray flares from \isrc{} detected by INTEGRAL are characterised by an average luminosity of $\sim\SI{3E35}{\erg\per\second}$, with stronger hard X-ray flares (by a factor of $\sim$3) being very rare.
In this context, the derived upper limits in Table~\ref{tab:agileobs1} are consistent with the typical outburst hard X-ray luminosity.
The lack of detections by INTEGRAL of hard X-ray flares simultaneous with the \gray flares detected by AGILE could be explained by the lack of observational coverage during the periods given in Table~\ref{tab:agileobs3} and/or the significantly lower INTEGRAL effective exposure time with respect to good quality observing conditions.
Alternative viable explanations include the possibility that hard X-ray and \gray flares are not produced by the same physical mechanism or that \asrc{} and \isrc{} are only closely aligned along the line of sight and are not physically associated.

\subsection{Orbital modulation and variability of \texorpdfstring{\asrc{}}{AGL~J1736-3250} in relation to \texorpdfstring{\isrc{}}{IGR~J17354-3255}}
\label{sect:discussion:flare_phases}
We investigated the potential connection between \isrc{} and \asrc{} by searching for an $\SI{8.4474}{\day}$ periodicity in the AGILE data, which would provide firm evidence that the two sources are physically connected.
The LS periodogram did not detect any sinusoidal periodic component, which could be potentially attributed to the signal having a non-sinusoidal periodicity, to an insufficient signal-to-noise ratio, or to the two sources not being physically connected.

To further explore the relationship, we computed the orbital phases of the $19$ \gray flares, assuming \asrc{} is the \gray counterpart of \isrc{}.
Approximately half of the flares occur near the apastron passages of \isrc{}.
Notably, $8$ of the $19$ flares are concentrated within a narrow orbital phase range $[0.4375-0.5000]$, corresponding to $1/16$ of the orbit.
The chance probability of this clustering is $P=\num{8.65E-6}$, equivalent to $\sim 4.3 \sigma$ Gaussian standard deviations.
If confirmed, this clustering near apastron would suggest a distinct origin for the \grays flares with respect to the hard X-ray flares, as the latter preferentially peak at periastron when the supergiant star is closest to the compact object and accretion is most efficient \citep[e.g. ][]{Sguera:2011, Ducci:2013}.
This hypothesis aligns with the system's estimated quasi-spherical orbit, with a low eccentricity of $\sim 0.1-0.2$, which would suppress strong orbital modulation effects.

The folded X-ray light curves in Fig.~\ref{fig:folded} from INTEGRAL ($18$-$60$ keV), \swift/XRT ($0.3$-$10$ keV) and \swift/BAT ($14-195$~keV) exhibit smooth orbital modulation, driven by numerous low-threshold flares that are undetectable individually, but contribute cumulatively to the folded emission.
In contrast, the AGILE light curve peaks coincide with the clustering of flares at phases $\sim0.4$ and $\sim0.8$, lacking the smooth modulation seen in X-rays.
This result may suggest a different origin for the hard X-ray and \gray emissions.

We calculated the variability index VI for \asrc{}, confirming variability at the $99\%$ confidence level.
The \gray variability is driven by random, non-periodic and bright flares, unlike the X-ray emission, which arises from regular, periodic accretion processes modulated by the orbital phase.

\subsection{Evidence of distinct emission mechanisms in \texorpdfstring{\asrc{}}{AGL~J1736-3250} and \texorpdfstring{\isrc{}}{IGR~J17354-3255}}
\label{sect:discussion:emission_scenarios}
Under the hypothesis that \asrc{} and \isrc{} represent the same source, our results (i.e. the INTEGRAL simultaneous upper limits during \gray flares, the clustering of \gray flares and the observed orbital modulations) suggest that the hard X-ray flares and the \gray flares are not produced by the same radiative process or at the same location.
This conclusion is also supported by the lack of phase alignment in the phase-folded light curves of the two bands (see Fig.~\ref{fig:folded}).

\begin{figure}
 \centering
 \includegraphics[width=\columnwidth,keepaspectratio]{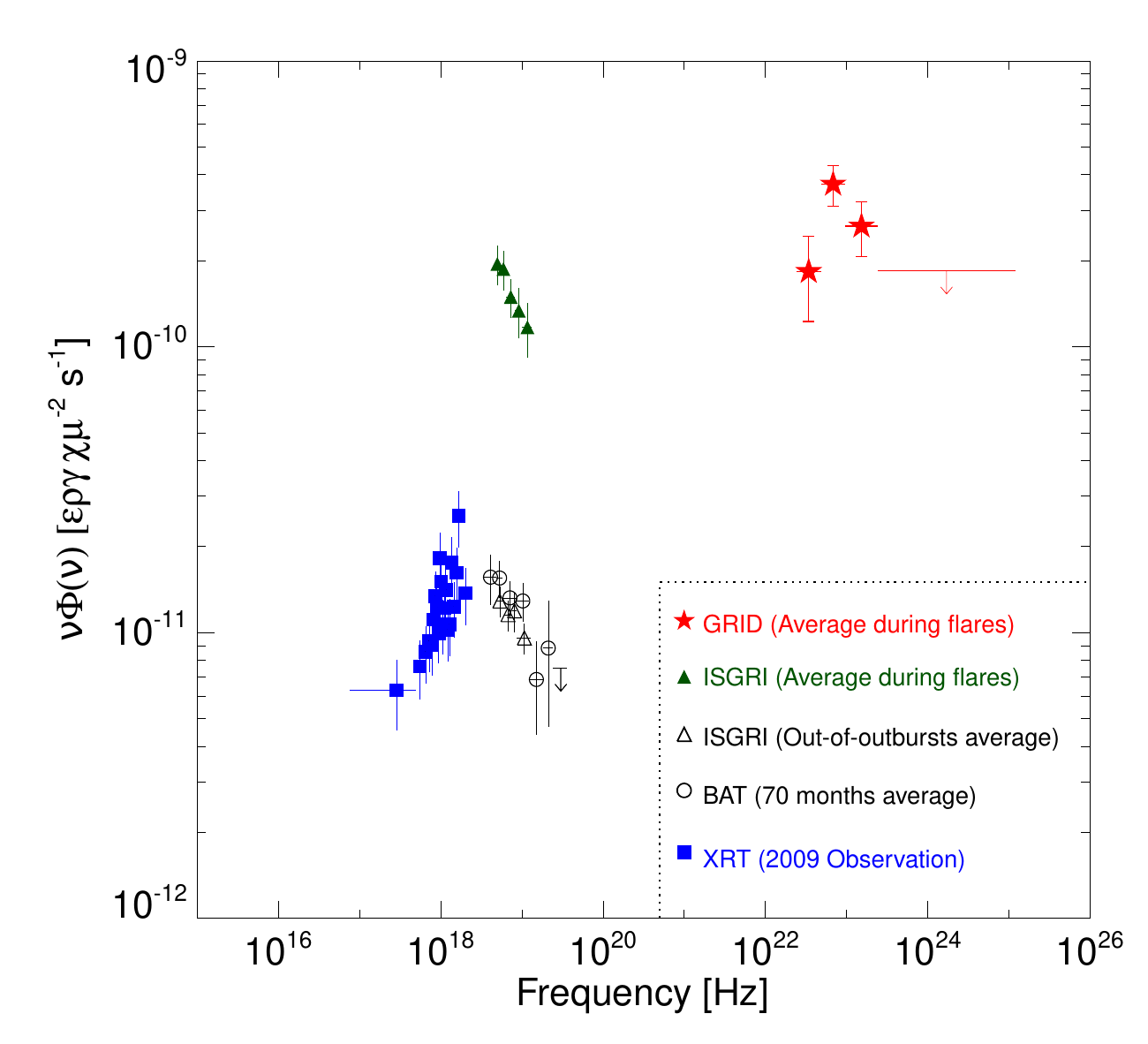}
 \caption{Spectral energy distribution of \isrc{} covering eight orders of magnitude in energy.
 Blue squares represent the \swift/XRT spectrum (corrected for absorption) accumulated during the 2009 observation (2009-04-17 01:08:59 to 2009-04-17 07:57:56).
 Black circles represent the \swift/BAT 70-months average spectrum.
 Black triangles represent the INTEGRAL/ISGRI out-of-outburst average spectrum covering the period Feb 2003  to Oct 2008.
 Green triangles represent the average INTEGRAL/ISGRI spectrum during flares, covering the period Apr 2003  to Mar 2008.
 Red stars represent the average AGILE/GRID spectrum of \asrc{} during flares.
 } 
 \label{fig:sed}
\end{figure}

In Fig.~\ref{fig:sed} we present the spectral energy distribution of \isrc{}, spanning from the X-ray to the HE \gray bands.
It includes the \swift/XRT spectrum from 2009 observations triggered by \citet{AGL2009Tel:Bulgarelli} and \citet{SWIFT2009Tel:Vercellone}, the IBIS/ISGRI average flare spectrum and the AGILE/GRID average \gray flaring emission from \asrc{}.
Additionally, it shows the out-of-outburst states for IBIS/ISGRI.

The slope of the average IBIS/ISGRI out-of-outburst spectrum (open black triangles) and during flares (filled green triangles) are comparable, differing in flux intensity by a factor of $\sim 10$.
This is compatible with the `clumpy wind' model, where variations in X-ray luminosity result from changes in the accretion rate of the compact object caused by the inhomogeneous and structured supergiant winds \citep{Sguera:2011, Bozzo_2017, Goossens_2018}.
During the out-of-outburst X-ray states, accretion continues but at lower rates, leading to reduced luminosities.
Complications to the clumpy wind model were highlighted by \citet{Ducci:2013}, who identified a dip in the folded \swift/XRT light curve around phase $\sim0.7$.
This dip, which cannot be attributed to lower wind accretion near apastron, might result from an eclipse or the onset of a `gated' mechanism, which would enact a transition to a less efficient accretion process.

A comparison of IBIS/ISGRI and AGILE/GRID data, although not simultaneous, reveals a clear spectral break between X-rays and $\gamma$-rays, indicating distinct emission mechanisms if \asrc{} is associated with \isrc{}.
If confirmed, the source would exhibit \gray flare luminosities exceeding the X-ray flares luminosities, akin to the behaviour of gamma-ray binaries, that emit predominantly above $\sim \SI{1}{\mega\electronvolt}$.

We note that the nature of the compact object hosted in \isrc{} is still unknown.
However, its broad band X-ray spectrum is typical of accreting neutron stars hosted in HMXBs. 
A possible explanation for the \gray emission from the source involves relativistic jets triggered by accretion of clumps from the supergiant's wind.
In fact, neutron stars might also be able to create relativistic jets.
\citet{2009ApJ...697.1194S} discussed the possibility that $\si{\mega\electronvolt}$ - $\si{\giga\electronvolt}$ \gray flares in SFXTs could result from transient jets powered by magnetic towers formed during clump accretion.
Magnetohydrodynamical simulations demonstrate that such structures can arise if accreting matter in the disk reaches $\sim 40$ gravitational radii \citep{kato_MagneticTowerJets_2004, kato_MagneticTowerJets_2007}.
Simulations indicate that magnetic loops connecting the NS and disk are twisted due to differential rotation between the two.
Twist injection from the disk expands the loops, creating a magnetic tower capable of collimating mass into bipolar jets.
Magnetic reconnection within the loops may then intermittently inject hot bubbles from the disk into the tower.
Particles within these bubbles could then be accelerated to HE, producing non-thermal emission.
Notably, this mechanism imposes a constraint on the strength of the NS magnetic field, as a strong magnetic field can extend the magnetosphere and inhibit accretion.
The formation of a magnetic tower and jets in NS require the magnetospheric radius, $R_M$, and the gravitational radius, $R_G$ being $R_M \lesssim 40 R_G$, to allow matter to reach the magnetic loops at the base of the tower \citep{kato_MagneticTowerJets_2004}.
For a NS with a mass of $M_\text{NS}=\SI{1.4}{\solarmass}$, we have $40 R_G \simeq \SI{1.7E7}{\centi\meter}$.
To satisfy the constraint on the magnetospheric radius, either very dense clumps ($\sim$~$10^{-4}$ to $10^{-5}$~$\si{\gram\,\centi\meter^{-3}}$) must accrete, which is inconsistent with the known properties of supergiant winds \citep[e.g. ][]{Fornasini_HMXBs_2023}, or the NS must have a low magnetic field ($B_\text{NS}\lesssim\SI{1.2E8}{\gauss}$).
Notably, NSs with such a low magnetic field, typical of old systems, have effectively been observed in some low-mass X-ray binaries (LMXBs), particularly Atoll- and Z-sources \citep{massi_magnetic_fields_jets_2008}, known to launch radio jets.
Conversely, known magnetic fields in HMXBs are typically $B_\text{NS} \sim \SI{E12}{\gauss}$, characteristic of young NSs.
If a low-magnetic-field NS was hosted in an HMXB, it would indicate the presence of magnetic field decay processes efficient enough to reduce $B_\text{NS}$ by four orders of magnitude within $\SI{E7}{\year}$, the typical age of HMXBs \citep{2011IAUS..275..309G, Li_SFXTfromAntimagnetars_2011}.
Magneto-hydrodynamical simulations suggest that accretion from an inhomogeneous wind of a supergiant star can enhance magnetic field decay at the NS surface, achieving the required decay rates if the accretion rate satisfies $\dot{M} \gtrsim \SI{E-10}{\solarmass\per\year}$ \citep{garcia_JetsSFXTs_2014}.
Assuming a spherical approximation for accretion in a wind-fed system, the corresponding X-ray luminosity is $\displaystyle L_X \approx 0.1 \dot{M} c^2$, yielding $L_X \gtrsim \SI{5.7E35}{\erg\per\second}$, consistent with observed values in SFXTs.

An important caveat of applying this scenario to \isrc{} concerns the energetic viability of the proposed $\gamma$-ray counterpart \asrc{}.
The $\gamma$-ray luminosity observed during the flares is significantly higher than the X-ray luminosity upper limits and, given the modest accretion rates inferred in this work, may even exceed the total accretion power available in the system, $L_{\rm acc} \sim G\,M\dot{M}/R_{\rm NS}$, where $M$, $\dot{M}$, and $R_{\rm NS}$ are the neutron star mass, accretion rate, and radius, respectively.
Furthermore, if the neutron star is indeed accreting, the magnetospheric radius $R_{\rm m}$ must be smaller than or comparable to the corotation radius $R_{\rm co}$, which limits the possibility of extracting rotational energy from the neutron star magnetosphere.
Therefore, although magnetospheric dissipation may in principle contribute to non-thermal emission, it is unlikely to account for the large energetic gap implied by the $\gamma$-ray flare luminosity.

These findings support the hypothesis that jet production in SFXTs, such as \isrc{}, might be possible, but remains theoretically challenging.
In this context, \citet{2009ApJ...697.1194S} proposed a model based on the accretion-jet framework to explain the \gray emission from the SFXT AX~J1841.0$-$0536, spatially associated with the MeV transient 3EG~J1837$-$0423 or the extended TeV source HESS~J1841$-$055 \citep{MAGIC_HESSJ1841-055_2020}.
In this model, the high-energy emission is produced by the cooling of relativistic electrons accelerated within a collimated jet launched by the NS.
Specifically, non-thermal X-ray and \gray emission could arise from synchrotron and inverse Compton processes, respectively \citep{2009ApJ...697.1194S}.
Thus, the formation of transient magnetic towers and jets could explain the \gray emission from \asrc{} and remains consistent with the hypothesis that this source is the counterpart of \isrc{}.

Another class of binary systems which exhibited transient gamma-ray emission is transitional millisecond pulsars \citep[tMSPS, ][]{papitto_tMSPs_2022}.
In particular, systems such as PSR~J1023+0038 exhibited a significant increase in gamma-ray luminosity when switching from a rotation-powered state to a disk-dominated state, an enhancement attributed to changes in magnetospheric configuration and the possible formation of an intra-binary shock \citep{stappers_tMSP_2014}.
Although SFXTs differ fundamentally from tMSPs in their donor type and accretion environment, it is worth noting that if a neutron star in a SFXT possesses an LMXB-like magnetic field, some analogous mechanisms (e.g. shocks or propeller-like episodes) could also contribute to high-energy emission.
While speculative, these parallels highlight that gamma-ray production in binary systems might be more diverse than traditionally assumed.

\section{Summary and conclusions}
\label{sect:conclusions}
We analysed the AGILE consolidated archive to search for \gray emission in the $0.1$~-~$10$~GeV range from the region of \isrc{}, selecting $1993$ days of observation with good exposure.
Our analysis detected $19$ flaring episodes on a $1$ day timescale.
The post-trial significance of repeated flare detection from the same sky region is $6.04\sigma$, confirming the AGILE detection of a transient source, designated \asrc{}.

Notably, eight of the flares occurred during `pointing mode' observations, despite the shorter observation period compared to `spinning mode'.
This suggests that continuous, uninterrupted observation is critical for detecting rapid transients such as \asrc{}.
Observations in spinning or survey modes are not suited to detect such transients, an important consideration for the design of future high-energy missions.

A periodic analysis of the AGILE data with LS periodogram on aperture photometry light curves revealed no sinusoidal periodicity in the \gray emission.
However, the variability index of the light curve folded using the known ephemeris of \isrc{} confirmed \asrc{} as a variable source at a $99\%$ confidence level.
Unlike the smooth, folded X-ray light curves of INTEGRAL and \swift{}, which exhibit modulated orbital emission peaking at periastron, the AGILE/GRID \gray light curve peaks during flaring episodes.
This suggests that the \gray variability is driven by stochastic, non-periodic, bright flares $-$ rather than the regular stacked emission of numerous low-intensity flares as in the X-ray band \citep{Sguera:2011}.

Approximately $(25-50)\%$ of the signal during most \gray flares was concentrated over a few hours, highlighting the high variability and low duty cycle of the source.
The spatial correlation and the very similar transient behaviour on short timescales measured in the X-rays and \grays suggests a physical association between \isrc{} and \asrc{}.
As reported in our work, simultaneous INTEGRAL observations during five \gray flares did not detect any hard X-ray activity.
This could be explained by the possibility that hard X-ray and \gray flares are produced by different physical mechanisms.
However, we note that the flare hard X-ray luminosity upper limits are not very stringent due to the poor INTEGRAL exposure times, and their values are compatible with the typical flare average hard X-ray luminosity measured in \isrc{}.
This opens up the possibility that hard X-ray flares and \gray flares might still be produced by the same physical mechanism.
However, we present additional evidence supporting the hypothesis of distinct emission mechanisms, i.e. an evident spectral break in the spectral energy distribution between X-rays and \grays of \asrc{} and \isrc{}, and the lack of phase alignment in the orbital modulation observed in the two energy bands.
In this context, we proposed a model based on the accretion-jet framework to explain the X-ray and the \gray emission from the SFXT \isrc{}.
The X-ray emission is likely driven by the accretion of clumps from the inhomogeneous supergiant wind onto the compact object.
This accretion process may also power \gray emission through the formation of a transient magnetic tower and jet, a possible scenario in HMXBs hosting low-magnetised NSs according to simulations \citep{kato_MagneticTowerJets_2007, 2009ApJ...697.1194S, garcia_JetsSFXTs_2014}.
However, the inferred $\gamma$-ray luminosity during the flares may be difficult to reconcile with the available accretion power and standard magnetospheric processes.
In this context, the absence of dedicated radio observations represents a key limitation, as it prevents us from testing the presence of jets or non-thermal outflows that could contribute to the emission.
This is especially relevant for HMXBs, as there are only a few systems hosting neutron stars that have shown radio emission and, thus, sub-GHz observations may be required to detect jet emission, if present \citep{vandenEijnden_HMXBradio_2025}.
The lack of direct measurements of the neutron star magnetic field further limits our ability to assess the viability of this scenario.
Similarly, a dedicated and thorough maximum likelihood analysis of \textit{Fermi}-LAT data during the epochs of the AGILE detections is still missing and could provide an important independent test of the $\gamma$-ray activity.
Therefore, while the empirical association between \asrc{} and \isrc{} is compelling, its physical interpretation remains an open issue that will require further multi-wavelength observations and more detailed theoretical investigations.

Our work provides evidence of a possible physical connection between \isrc{} and \asrc{}, although this association is not yet firmly established.
These findings reinforce the potential of SFXTs as sources of HE emission.
Several studies have suggested that other HMXBs and SFXTs could also be candidate counterparts for unidentified transient HE and VHE sources on the Galactic plane \citep{2009ApJ...697.1194S, 2009arXiv0902.0245S, Sguera:2011, 2012ApJ...748...86O, MunarAdrover:2016:MWC656, Harvey_HMXBsFermiLAT_2022_10.1093/mnras/stac375}.
However, detecting such fast transients with current instruments is challenging due to the emission likely being composed of unpredictable and short flares.
Furthermore, a firm confirmation of these sources require coordinated multi-wavelength studies to properly characterise the emission mechanisms and the physical properties of the systems.

Future advancements in \gray observatories facilities, such as upcoming COSI satellite mission observing at $\sim\SI{1}{\mega\electronvolt}$ \citep{tomsick_COSI_2023} and the Cherenkov Telescope Array Observatory (CTAO) in the VHE band, could provide significant improvements in sensitivity and survey speed.
In particular, the CTAO will be particularly suited for fast transient astronomy thanks to its improved VHE sensitivity, more than one order of magnitude better than current VHE facilities \citep{carosi_ctatransient_2021, ctac_science_2019, zanin_cta_2022, 2025_CTAO_GalacticTransients}, along with its capacity to swiftly react to external alerts on astrophysical transients by triggering ToO observations.
The CTAO will be able to swiftly repoint its telescopes and perform real-time analyses using its Science Alert Generation system \citep{di_piano_detection_2021, caroff_real_2023, bulgarelli_SAGproceedingsSPIE_2024}, which is part of the Array Control and Data Acquisition of CTAO \citep{oya_acadaspie_2024}.
These developments could enable the detection (or the non-detection) of SFXTs at HE and VHE energies, offering critical insights into extreme physical mechanisms and opening up an unexplored energy window.
Such breakthroughs would have profound implications for understanding transient HE astrophysical phenomena and processes capable of emitting non-thermal radiation on very short timescales.

\begin{acknowledgements}
The AGILE Mission is funded by the Italian Space Agency (ASI) with scientific and programmatic participation by the Italian National Institute for Astrophysics (INAF) and the Italian National Institute for Nuclear Physics (INFN). The investigation is supported by the ASI-INAF agreement ASI-I/028/12/0 and subsequent addenda (up to ASI-I/028/12/7). We thank the ASI management for unfailing support during AGILE operations. We acknowledge the effort of ASI and industry personnel in operating the ASI ground station in Malindi (Kenya), and the data processing done at the ASI/SSDC in Rome: the success of AGILE scientific operations depends on the effectiveness of the data flow from Kenya to SSDC and the data analysis and software management.
This research was funded by the \textquotedblleft Programma di Ricerca Fondamentale INAF 2023\textquotedblright\ (PR).
This work is based also on observation with INTEGRAL, an ESA Project with instruments and Science Data Center founded by ESA members states (especially France, Germany, Denmark, Italy,  Spain and Switzerland) and the participation of Russia and the USA.
Angela Bazzano acknowledges support via ASI/INAF agreement 2019-35-HH0.\\

\emph{Facilities:} AGILE(GRID), INTEGRAL(IBIS), Swift(BAT, XRT, and UVOT).\\

\emph{Software:} Astropy \citep{astropy:2013,astropy:2018,astropy:2022}, Agilepy \citep{bulgarelli_agilepy_2022}.
The AGILE data analysed in this work, along with the derived high-level maps and analysis notebooks have been made publicly available\footnote{\url{https://github.com/AGILESCIENCE/AGILE_detection_AGLJ1736-3250.git}} to support reproducibility \citep{bulgarelli_repositorydata_2026}.
\end{acknowledgements}


\bibliographystyle{aa}
\bibliography{bibliography}



\begin{appendix}

\section{Evaluation of the p-value and post-trial statistical significance}
\label{sect:appendix2}
We performed a simulation of the \isrc{} Galactic region, including the steady sources from the 2AGL catalogue, along with the Galactic diffuse and isotropic background emissions, in the source model.
We assumed a one-day mean exposure level, using the average value of the exposure in `pointing' and `spinning' modes.
The primary aim of the simulation was to evaluate the $TS$ and the $p$-value distributions of the AGILE maximum likelihood estimator in the target region under the null hypothesis that no \gray source coincident with \isrc{} is present in the AGILE data.
During the $TS$ evaluation, all parameters of the 2AGL sources and the diffuse emission coefficients were kept fixed.
We adopted a Galactic background coefficient of $g_\text{gal}=0.6$ and isotropic coefficient of $g_\text{iso}=8$ \citep{Bulgarelli:2012ds}.

\begin{figure}[ht]
 \centering
 \includegraphics[width=\columnwidth,keepaspectratio]{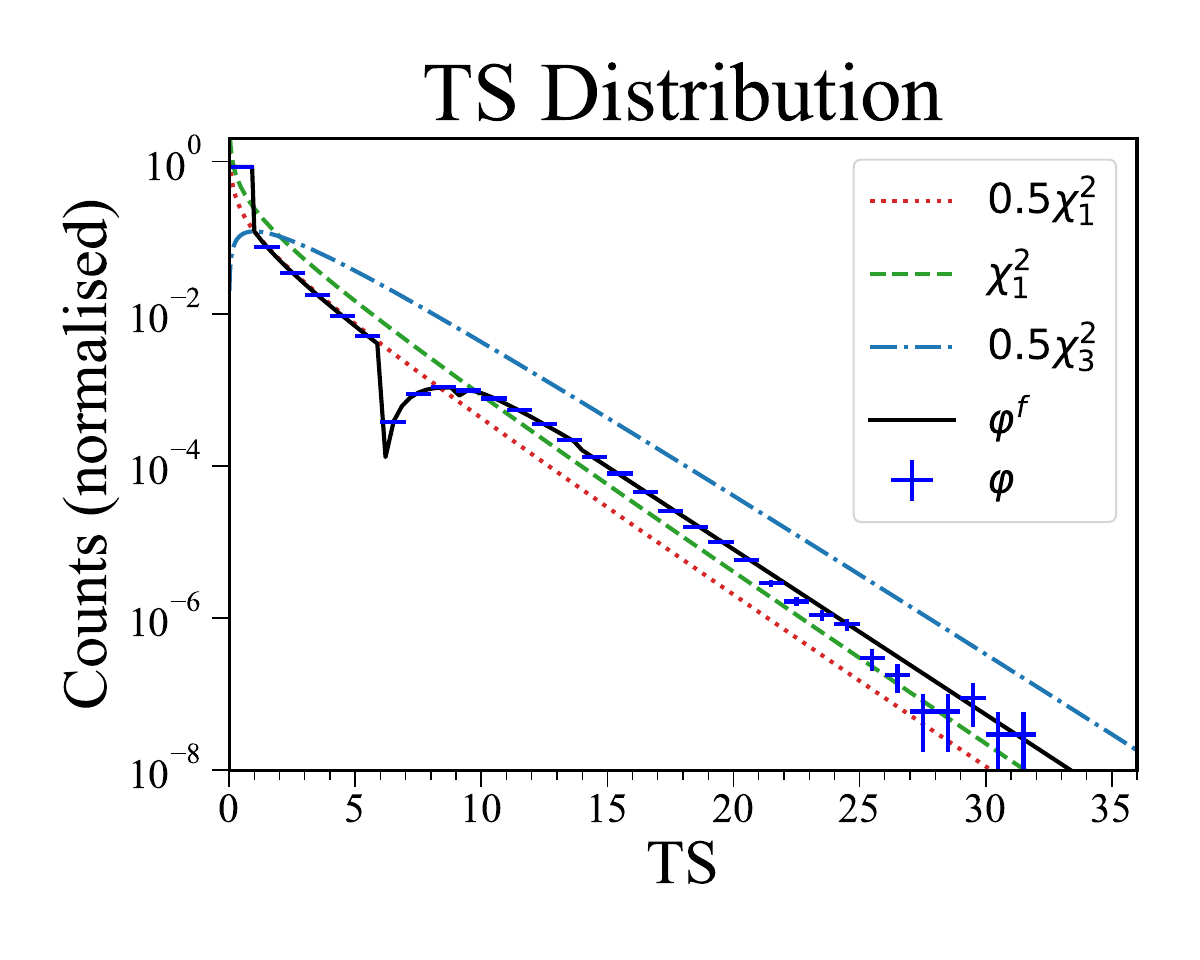}
 \caption{$TS$ distribution of the \isrc{} sky region.
 Blue crosses represent the calculated distribution; the black line shows the best fit.
 The dotted red, dashed green, and dash-dotted cyan lines correspond to the $0.5 \chi^2_1$, $\chi^2_1$, and $0.5 \chi^2_3$ theoretical distributions, respectively.
 }
 \label{fig:TSdistModel13r09_TS}
\end{figure}

We determined the $p$-value distribution, $p(h)$, under the aforementioned null hypothesis of the likelihood test, which was the same adopted for the AGILE data analysis in Sect.~\ref{sect:agile_results}.
The $p-$value $p(h)$ at a given $TS$ threshold, $h$, is defined as
\begin{equation}
   p = p(h) = P(TS \geq h) = \int_{h}^{+\infty} \varphi (x) dx,
   \label{eq_A}
\end{equation}
where $\varphi$ is the $TS$ distribution, illustrated in Fig.~\ref{fig:TSdistModel13r09_TS}.

\citet{Bulgarelli:2012ds} provided in Eq.~(5) an analytical expression, $\varphi^f$, to approximate $\varphi$:
\begin{equation}
    \label{eq_phi_approximation}
    \varphi^f (TS) = \begin{cases}
      \delta                                  & \text{if}\ TS<1 \\
      \eta_1 \, \chi^2_{N_1}(TS)              & \text{if}\ 1 \leq TS \leq t_\text{lcl} \\
      \eta_2 \, \chi^2_{N_2}(TS-t_\text{lcl}) & \text{if}\ t_\text{lcl} \leq TS \leq t_\text{ICL} \\
      \eta_3 \, \chi^2_{N_3}(TS-t_\text{ICL})              & \text{if}\ t_\text{ICL} \leq TS \leq T_1 \\
      \eta_4 \, \chi^2_{N_4}(TS-T_1)              & \text{if}\ TS \geq T_1 \\
    \end{cases}.
\end{equation}
Equation~\eqref{eq_phi_approximation} approximates the expected behaviour of the maximum likelihood analysis.
This behaviour transitions from a source location algorithm with many free parameters near the threshold where the location contour is large (identified by $t_\text{lcl}$) to an analysis resembling a fixed-position analysis at high $TS$ values, where the location contour is small (identified by $t_\text{ICL}$ and $T_1$).
We adopted $t_\text{lcl}=6$, $t_\text{ICL}=9$, $T_1=14$, typical values for the analysis of a Galactic region \citep{Bulgarelli:2012ds}, and we report the best fit of the parameters in Table~\ref{tab:varphi_parameters}.

\begin{figure}[ht]
 \centering
 \includegraphics[width=\columnwidth,keepaspectratio]{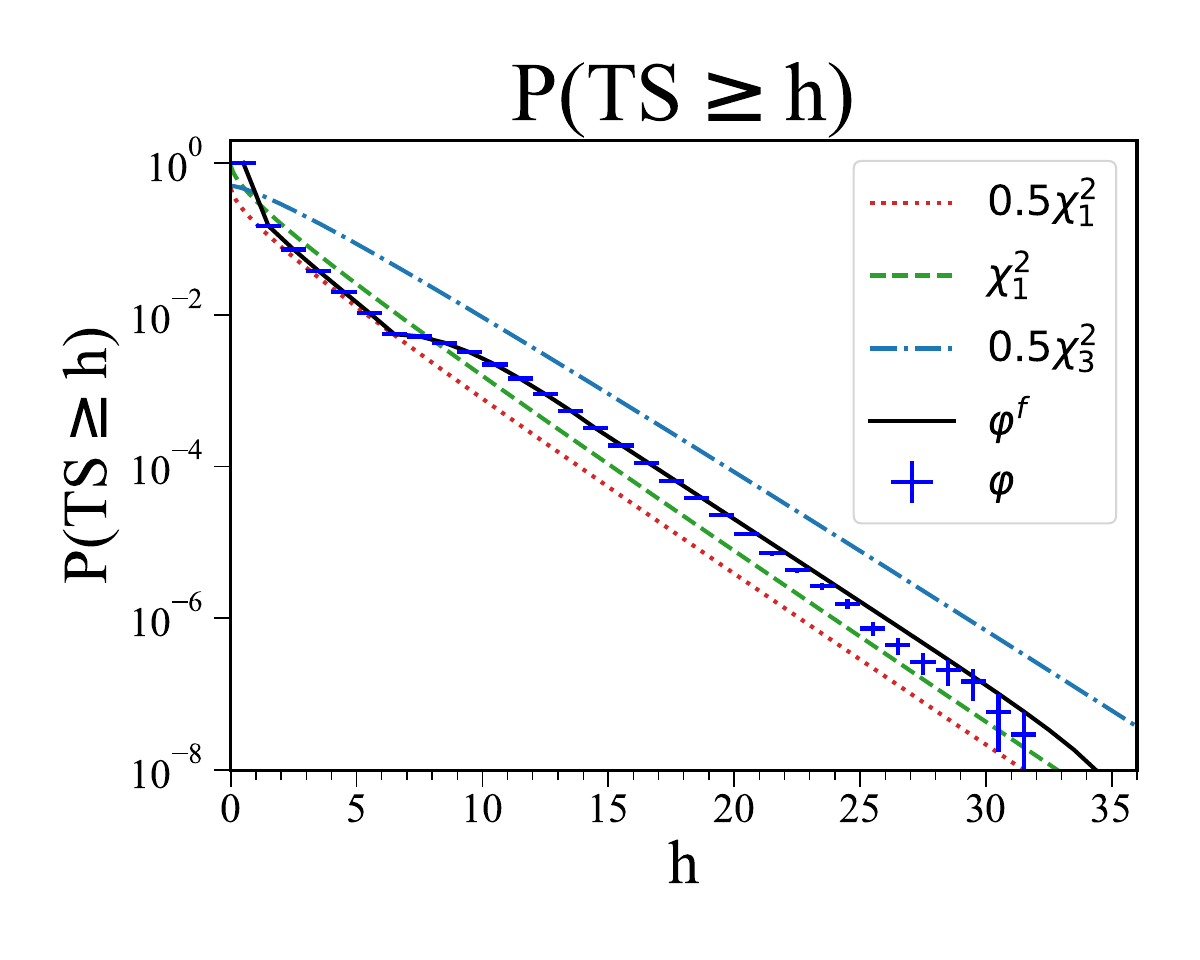}
 \caption{$p$-value distribution of the \isrc{} sky region.
 We employ the same colour coding of Fig.~\ref{fig:TSdistModel13r09_TS}.
 }
 \label{fig:TSdistModel13r09_h}
\end{figure}

We show in Fig.~\ref{fig:TSdistModel13r09_h} the $TS$ distribution under the null hypothesis when the flux and position of the \asrc{} source are allowed to vary, subject to the criterion that the source lies within the confidence contour level.
Using this $TS$ distribution, we selected the threshold to detect \gray flares in AGILE/GRID data.
We fixed the threshold at $\sqrt{TS} = 3.3$, corresponding to $\simeq 3 \sigma$, while $\sqrt{TS}\simeq 5.4$ corresponds to $\simeq 5\sigma$.

Typically, the post-trial significance of a $TS$ evaluation is computed by treating each trial as a single independent occurrence \citep[e.g. ][]{LiMa_Paper_1983}, without considering the history of repeated occurrences.
We accounted for the sky region's history by computing the probability of `repeated flaring episodes' from the same sky position.
The chance probability of having $k$ or more detections over $N$ trials with $TS \ge h$ is given by
\begin{equation}
 P(N, X \ge k)
 = 1 - \sum_{j=0}^{k-1}
 \left(\begin{array}{c}N\\j\end{array}\right) p^j (1-p)^{N-j},
\label{eqn-ICL}
\end{equation}
where $h$ is the minimum level of $TS$ selected to include a temporal bin in the light curve, while $p = p(h)$ is the corresponding $p$-value.

\begin{table}[ht!]
 \centering
 \caption{Best-fit parameters of the analytical approximation $\varphi^f$ of the $TS$ distribution, defined in Eq.~\eqref{eq_phi_approximation}.
 }
 \label{tab:varphi_parameters}
 \begin{tabular}{cc}
  \hline
  Parameter & Value \\
  \hline
  $\delta $ & $ 0.8522\pm0.0002$\\
  $\eta_1 $ & $ 0.525\pm0.001$\\
  $N_1$     & $ 0.924\pm0.002$\\
  $\eta_2 $ & $ (6.52\pm0.04)\cdot 10^{-3}$\\
  $N_2$     & $ 4.58\pm0.01$\\
  $\eta_3 $ & $ (3.20\pm0.01)\cdot 10^{-3}$\\
  $N_3$     & $ 2.491\pm0.008$\\
  $\eta_4 $ & $ (3.26\pm0.002) \cdot 10^{-4}$\\
  $N_4$     & $ 2.000\pm0.002$\\
  \hline
 \end{tabular}
\end{table}

\end{appendix}
\end{document}